\documentclass[%
 reprint,
 amsmath,amssymb,
 aps,
 prd,
floatfix,
twocolumn
]{revtex4-2}

\usepackage{graphicx}
\usepackage{dcolumn}
\usepackage{bm}

\usepackage{xcolor}
\usepackage{graphicx}
\usepackage{epsfig}
\usepackage{slashed}
\usepackage{tabu,multirow}
\usepackage[unicode=true,bookmarks=false,breaklinks=false,pdfborder={0 0 1},colorlinks=true]{hyperref}
\hypersetup{
citecolor=blue,linkcolor=blue,urlcolor=blue}
\begin{document}


\title{
Isospectrality and configurational entropy as testing tools for bottom-up AdS/QCD }

\author{Miguel Angel Martin Contreras}
\email{miguelangel.martin@usc.edu.cn}
\affiliation{
 School of Nuclear Science  and Technology\\
 University of South China\\
 Hengyang, China\\
 No 28, West Changsheng Road, Hengyang City, Hunan Province, China.
}


\author{Alfredo Vega}%
 \email{alfredo.vega@uv.cl}
\affiliation{%
 Instituto de F\'isica y Astronom\'ia, \\
 Universidad de Valpara\'iso,\\
 A. Gran Breta\~na 1111, Valpara\'iso, Chile
}

\author{Saulo Diles}
\email{smdiles@ufpa.br}
\affiliation{Campus Salin\'opolis,\\ Universidade Federal do Par\'a,\\
68721-000, Salin\'opolis, Par\'a, Brazil}
\affiliation{Unidade Acadêmica de Física,\\ Univ. Federal de Campina Grande, R. Aprígio Veloso, 58429-900 - Campina Grande}

\date{\today}

\begin{abstract}
This work discusses the connection between isospectrality and configurational entropy in holographic bottom-up models. We analyze the effect of monoparametric isospectral transformation in holographic decay constants and configurational entropy for a set of softwall-like models at zero temperature.
We conclude that the isospectral parameter $\lambda$ defines a window of possible holographic models suitable for describing spectroscopy.   

\end{abstract}
\maketitle
\section{\label{intro}Introduction}
Calculations of hadronic spectra are a hard problem in QCD, which found a friendly approach in the context of holographical models, especially the bottom-up approach, where one tries to find a gravity that captures QCD at the conformal boundary. The problem of holographic hadronic masses is generally reduced to calculating the masses of states as the eigenvalue spectrum of a given holographic potential $V(z)$. The holographic potential carries information about the dilaton and the geometry used. 
At the extrema points, i.e., $z\to0$ and $z\to \infty$, the Poincarè warp factor and the dilaton control the holographic potentials. The richness of the dilaton technology appears in the intermediate region, leading to improved decay constants or form factors \cite{Brodsky:2007hb, Abidin:2009hr, MartinContreras:2021yfz, FolcoCapossoli:2020pks, Braga:2015jca, MartinContreras:2019kah}. This enhancement could be done using SUSY quantum mechanical tools, where it is possible to transform a potential into another one with the same eigenvalue spectrum in terms of a real value constant $\lambda$, i.e., an isospectral transformation.

In AdS/QCD terms, if we have a well-known hadron spectrum, it is suitable to be modeled by a dilaton. Therefore, we can match this dilaton with an element in a bigger monoparametric isospectral class \cite{Vega:2016gip}.  We start with a hadronic mass spectrum and find a dilaton that produces a holographic potential fitting such a mass spectrum. Once we have a potential, we look for its associated isospectral class. All potentials in this class are equally good for modeling the original hadronic mass spectrum, each having its associated dilaton. Within this approach, the quantity $\lambda$ arises as a free natural parameter of a given bottom-up holographic model that describes the spectral characteristics of any hadron. Here, we prove for vector mesons that the electromagnetic decay constant of the ground state is sensible on $\lambda$. In other words, an isospectral class of holographic potentials describing vector meson masses fits the ground state decay constant.

In Ref. \cite{Afonin:2018era}, authors have discussed isospectrality concerning the confinement/deconfinement phase transition in the glueball context. Authors claim critical deconfinement temperatures should be independent of the isospectral parameter $\lambda$. This result was demonstrated for scalar glueballs. However, we expect this temperature to strongly depend on the isospectrality in the vector meson case since thermal stability depends on decay constants. 

Since we have many isospectral holographic potentials, which one is chosen to model hadron spectra? A form to solve this question is by turning our eyes to the stability of the solutions. We will consider \emph{configurational entropy} (CE) as a stability criterion for hadrons at zero temperature \cite{MartinContreras:2023oqs} when confinement can be understood as a localization effect in space. In this sense, an AdS/QCD model is suitable for describing hadron sp if, for some $\lambda$, it fits the decay constant of the ground state \emph{and} CE increases with excitation level.   

This work is organized as follows. Section \ref{isospectrality} introduces the isospectrality of SUSY quantum mechanics. Section \ref{hadrons} briefly discusses how hadrons are conceived in AdS/QCD models. In Section \ref{DCE}, we briefly discuss the ideas behind configurational entropy. Section \ref{testing} summarizes the bottom-up models we will test with CE and isospectrality. Section \ref{bottom-up iso} is devoted to the isospectrality in the bottom-up conceptual frame. We present our results in Section \ref{Isos and DCE}. And finally, in Section \ref{conclusions}, we present our conclusions.  

\section{Isospectrality}\label{isospectrality}
\subsection{General idea}
In physics, isospectrality deals with two Hamiltonians having the same energy spectrum. This discussion was developed up to the early 1980s in the context of the Darboux transform, a result of second-order differential ordinary equations in the nineteenth century \cite{darboux1882proposition, abraham1980changes, mielnik1984factorization}. However, at the beginning of the 80s decade, isospectrality was found in supersymmetric quantum mechanics as a new formulation, which is now one of the most popular \cite{Witten:1981nf}.
The methods used in this manuscript follow the ideas developed in \cite{Cooper:1994eh}, which is standard in supersymmetric quantum mechanics.
In Ref. \cite{Vega:2016gip}, the supersymmetric quantum mechanical procedure to produce a monoparametric family of isospectral potentials has been considered in the context of bottom-up holography. In the following, we provide the details of this procedure.  

Our starting point is considering a potential $V_1(z)$, with an associated set of eigenvalues and eigenfunctions $\{\lambda_n,\,\phi_n(z)\}$. The idea of isospectrality can be motivated by the following query: Does any other potential, $\hat{V}(z)$, share the same eigenvalues $\lambda_n$ with $V_1(z)$? The answer emerges in terms of the so-called superpotential $W$.

For any given potential $V_1(z)$, there is a superpotential $W(z)$ such that $V_1(z) = W^2-W'$. The superpotential is derived from the ground state $\phi_0(z)$ of $V_1$ by $W=-\frac{d}{dz} \log \phi_0$ and generates a superpartner:  $V_2(z) = W^2+W'$. The family of strictly isospectral potentials of $V_1$ emerges when we look for superpotentials $\hat{W}$ that generate the same superpartner $V_2$. We forget about $V_2$ and keep only the isospectral family of $V_1$. 

Suppose that $W$ and $\hat{W}$ are the superpotentials generating $V_2(z)$ and $\hat{V}_2(z)$:

\begin{eqnarray}
    V_2(z)&=&W^2(z)+W'(z),\\
    \hat{V}_2(z)&=&\hat{W}^2(z)+\hat{W}'(z).
\end{eqnarray}

The general superpotential could be written as the following expression.

\begin{equation}
    \hat{W}(z)=W(z)+f(z).
\end{equation}

Thus, for the potential $\hat{V}_2(z)$ we have

\begin{equation}
    \hat{V}_2(z)=W^2+f^2+2\,W\,f+W'+f'.
\end{equation}

Therefore, to ensure that the transformation is isospectral, we must hold the following condition 

\begin{equation}
     f^2+2\,W\,f+f'=0.
\end{equation}

 The solution to this equation defines the \emph{one parameter family of isospectral superpotentials}

 \begin{equation}
     \hat{W}(z)=W(z)+\frac{d}{d\,z}\log\left[I(z)+\lambda\right]
 \end{equation}

\noindent where $\lambda \in (0,\infty)$ is an integration constant. Alternatively, one can define $\lambda$ in the range $(-\infty,\,-1)$.  The function $I(z)$ is defined in terms of the ground state of the potential $V_1(z)$, $\phi_0(z)$, as

\begin{equation}\label{Darboux}
I(z)=\int_0^z{dz'\,\phi_0^2(z')}.
\end{equation}

This expression is called \emph{isospectral transformation} \cite{Cooper:1994eh}. Then, the monoparametric family of potentials that are isospectral to $V_1(z)$ is

\begin{equation}\label{isopotential}   \hat{V}_\lambda(z)\equiv\hat{W}^2(z)-\hat{W}'(z)=V_1(z)-\frac{d^2}{d\,z^2}\log\,\left[I(z)+\lambda\right].
 \end{equation}
 
For a given $\lambda$, the potential $\hat{V}_\lambda(z)$ has the same spectrum as its superpartner $V_2(z)$, apart from the ground state. To complete the spectrum of $\hat{V}(z)$, we reintroduce  the lost ground state, which is given by \cite{Khare:1988is, Cooper:1994eh}:

\begin{equation}\label{isogst}
    \phi_{0,\lambda}(z) \equiv \frac{\phi_0(z)}{I(z)+\lambda}.
\end{equation}

In the next sections, we will test this expression in the context of bottom-up models. Our strategy will be constructing the associated monoparametric isospectral family for a well-known holographic confining potential.

\section{Holographic description of hadrons in bottom-up models} \label{hadrons}
\subsection{Bottom-up AdS/QCD in a scratch}

Let us focus on the holographic description of hadronic states. We shall consider as our background space the standard Poincarè patch defined by

\begin{equation}
    dS^2=\frac{R^2}{z^2}\left[dz^2+\eta_{\mu\nu}\,dx^\mu\,dx^\nu\right],
\end{equation}

\noindent where $\eta_{\mu\nu}$ is the Minkowski metric tensor in four dimensions and $R$ is the AdS curvature radius. 

According to the standard AdS/CFT prescription, the hadronic identity is established by a single quantity, the dimension of the operator that creates hadrons at the boundary, which enters the bulk action \emph{via} the bulk mass $M_5$. This matching is a consequence of the so-called \emph{field/operator duality} \cite{Aharony:1999ti}.
 
Integer spin hadrons can be described by a single bulk action in the flat approach, written as 

\begin{multline}\label{bulk-action}
    I=\frac{1}{2\,\mathcal{K}}\int{d^5x\,\sqrt{-g}\,e^{-\Phi(z)}\left[\nabla_{m}\,\phi^{m_1\dots \,m_S}\,\nabla^m\,\phi_{m_1\dots\, m_S}\right.}\\
    \left.+M_5^2\,\phi^{m_1\dots\,m_S}\,\phi_{m_1\dots\,m_S}\right],
\end{multline}

\noindent where $\phi_{m_1\dots\,m_S}$ is a bulk field dual to hadronic operators at the boundary, and $\mathcal{K}=(2\,\pi)^2$ is a constant that fixes units, relevant for the calculation of decay constants. We have written a generic dilaton field to address some AdS/QCD models based on static dilaton fields $\Phi(z)$. 

From this action, we can derive the equation of motion for the $S$-form bulk field $\phi$ as follows:

\begin{multline}
    \frac{1}{\sqrt{-g}}\,\partial_m\left[\sqrt{-g}\,e^{-\Phi(z)}\,g^{m\,n}\,g^{m_1\,m'_1}\dots\,g^{m_S\,m'_S} \nabla_n\,\phi_{m'_1\dots\,m'_S}\right]\\
    -M_5^2\,e^{-\Phi(z)}\,g^{m_1\,m'_1}\dots\,g^{m_S\,m'_S}\,\phi_{m'_1\dots\,m'_S}=0.
\end{multline}

To obtain these equations of motion, we have imposed the transverse gauge $\nabla_{\mu_1}\phi^{\mu_1\ldots\,\mu_s}=0$.

After Fourier transforming and redefining the bulk field as $\phi_{\mu_1\ldots\mu_S}\left(z,q\right)=\tilde{\phi}_{\mu_1\ldots\mu_S}\left(q\right)\,\psi\left(z,q\right)$, we arrive to the Sturm-Liouville equation for the bulk field $\psi(z,q)$:

\begin{multline}\label{sturm-liouville}
    \partial_z\left[e^{-B(z)}\,\partial_z\,\psi(z)\right]+\left(-q^2\right)\,e^{-B(z)}\,\psi(z,q)\\
    -\frac{M_5^2\,R^2}{z^2}\,e^{-B(z)}\psi(z,q)=0
\end{multline}

\noindent where we have defined $B(z)=\Phi(z)+\beta\,\log\left(\frac{R}{z}\right)$ and $-q^2=M_n^2$ is the on-shell condition and  $M_5^2R^2=0$ for vector mesons and $M_5^2\, R^2=-3$ for scalar mesons. Notice that $\beta$ is a factor that carries spin information since $\beta=-3+2\, S$. For scalar fields, $\beta=-3$, and for vector fields $\beta=-1$.

The hadronic spectrum, i.e., the associated holographic Regge trajectory, comes from transforming the Sturm-Liouville equation into a Schrödinger-like one. To do so, we perform a Boguliobov transformation $\psi(z)=e^{\frac{1}{2}B(z)}\,u(z)$:

\begin{equation}\label{schrodinger}
    -u''+V(z)\,u=M_n^2\,u,
\end{equation}

\noindent where the holographic potential is defined as 

\begin{multline}\label{holographic-pot}
   V(z)=\frac{M_5^2\,R^2}{z^2}+\frac{1}{4}\left[-\frac{\beta}{z}+\Phi'(z)\right]^2\\
   -\frac{1}{2}\left[\frac{\beta}{z^2}+\Phi''(z)\right].
\end{multline}

The eigenvalues of this potential define the hadronic spectrum, and the eigenstates are dual to normalizable hadronic modes. 

With the hadronic spectrum, another significant quantity we can deduce from this holographic model is the hadronic decay constants $f_n$, given in energy units. Decay constants are quantities measuring the probability of transitioning into the vacuum of a given hadronic state. In the language of OPE expansions, the decay constants appear as residues of the multipole expansion of the 2-point function $\Pi(-q^2)$. Holographically, it is defined as \cite{MartinContreras:2019kah}

\begin{equation}
  f_n^2=\frac{\left(\Delta-S\right)^2}{M_n^2\,\mathcal{K}}\,\lim_{z\to\varepsilon} \,e^{-2\,\Phi(z)-\left(\beta-1\right)\,A(z)}\left|\frac{u_n(z,q)}{z}\right|^2, 
\end{equation}

\noindent where $\varepsilon\to 0$ defines the AdS conformal boundary locus. 

We have not discussed the role of dilaton in this scenario until now. The dilaton is responsible for inducing confinement in the model. A direct consequence of confinement is the emergence of hadronic bounded states. In holographic terms, including a dilaton field makes free bulk fields in AdS become normalizable states. These normalizable states are dual to hadrons at the boundary. 

 \subsection{Dilaton engineering}
Isospectral tools define a new holographic potential in the parameter $\lambda$. At this point, it is worth asking what static dilatons are associated with this family of isospectral potentials \cite{Vega:2016gip}. 

To explore this possibility, we will connect these monoparametric potentials $\hat{V}_\lambda(z)$ with the dilaton as follows

\begin{equation}\label{dialton-eng}
\hat{V}_\lambda(z)=\frac{M_5^2\,R^2}{z^2}-\frac{1}{2}\left[\frac{\beta}{z}+\tilde{\Phi}'(z,\lambda)\right]+\frac{1}{4}\left[-\frac{\beta}{z^2}-\tilde{\Phi}''(z,\lambda)\right]^2
.  
\end{equation}

Thus, given an isospectral potential $\hat{V}_\lambda(z)$, we can construct the associated dilaton field $\tilde{\Phi}(z,\lambda)$. This $\tilde{\Phi}$ field is a deformation of the standard dilaton $\Phi$, used in the action density \eqref{bulk-action}, which generates the potential $\hat{V}_\lambda(z)$ in a holographic sense. 

Methodologically, dilaton engineering flows directly from bottom-up modeling: we always start from the given spectrum at the boundary (coming from experimental phenomenology). Thus, this spectrum fixes the choice of geometry or dilaton. In our case, we have a spectrum that defines a monoparametric family of potentials. Therefore, we want to compute the static dilaton associated with each element in the isospectral family. 

\section{Configurational Entropy}\label{DCE} 
Configurational entropy (CE) can be easily defined as the different forms (in terms of microstates) in which a given macrostate can be organized. Thus, a larger CE  means a higher number of possible microstate arrangements. In thermodynamical terms, this entropy is associated with the work done by a system without any exchange of energy changing spatial configuration. Configurational entropy comes from the so-called Shannon entropy, defined in information theory \cite{Gleiser:2011di}. 

For a discrete variable, configurational entropy  with probabilities $p_n$ is defined from the Shannon entropy as follows \cite{Gleiser:2012tu,Bernardini:2016qit,Braga:2016wzx}

\begin{equation}
    S_C=-\sum_n\,p_n\,\text{log}\,p_n.
\end{equation}

In the case of continuous variables, we have the \emph{differential configurational entropy} (DCE) defined as 

\begin{equation}\label{DCE-def}
S_C\left[f\right]=-\int{d^d\,k\,\tilde{f}\left(k\right)\,\text{log}\,\tilde{f}\left(k\right)},    
\end{equation}

\noindent where $\tilde{f}\left(k\right)=f\left(k\right)/f\left(k\right)_\text{Max}$ defines the \emph{modal fraction}, $f(k)_\text{Max}$ is the maximum value assumed by $f(k)$. Also, we have $f(k)\in\, L^2\left(\mathbb{R}^2\right)$, i.e. the square-integrable space of functions on the plane. This ensures that $f(k)$ has a defined Fourier transform. Usually, this $f(k)$ function is associated with the energy density in the momentum space, $\rho(k)$.

In the AdS/CFT context, the holographic approach to configurational entropy in bottom-up and top-down AdS/QCD models was made in \cite{Bernardini:2016hvx}. For the hadronic states, it was introduced in \cite{Bernardini:2016qit, Braga:2017fsb, Colangelo:2018mrt, Ferreira:2019nkz, Ferreira:2020iry, daRocha:2021ntm} and references therein. In the context of heavy quarkonium stability, DCE was used as a tool to explore thermal behavior in a colored medium \cite{Braga:2018fyc}, in the presence of magnetic fields \cite{Braga:2020hhs} or at finite density \cite{Braga:2020myi}. In \cite{Braga:2020opg}, DCE addressed the holographic deconfinement phase transition in bottom-up AdS/QCD. Recently, CE was used to discuss holographic stability in light nuclides in \cite{MartinContreras:2022lxl} and non-$q\bar{q}$ states \cite{MartinContreras:2023oqs}.

The holographic dictionary maps the information encoded in the spatial configuration of the boundary particle into the holographic configuration of the dual bulk field. In this sense, the information associated with the arrangement of the constituents within the hadron is encoded in the energy density of the bulk field, i.e.,  $\rho(z) \equiv T_{00}(z)$, with $T_{mn}$ defined as the on-shell energy-momentum tensor. 

As it was described in \cite{MartinContreras:2022lxl}, the standard procedure to compute CE starts from the bulk action \eqref{bulk-action} by calculating the energy-momentum tensor $T_{mn}$

\begin{equation}\label{Energy-momentum}
T_{mn}=\frac{2}{\sqrt{-g}}\,\frac{\partial\left[\sqrt{-g}\,\mathcal{L}_\text{Hadron}\right]}{\partial\,g^{mn}}.  
\end{equation}

We then obtain the Schrödinger modes calculated from eqn. \eqref{schrodinger} we transform back to the Sturm-Liouville form and then compute the on-shell energy density as

\begin{multline}\label{energy-density}
 \rho(z)\equiv T_{00}=\frac{e^{-B\left(z\right)}}{2}\left(\frac{z}{R}\right)^3\times\\
\left\{ \left[\frac{1}{\mathcal{K}^2}\left(M_n^2\,\psi_n^2+\psi_n'^2\right)-\frac{M_5^2\,R^2}{z^2}\psi_n^2\right] \right\}\,\Omega,
\end{multline}

\noindent where $\Omega$ is a factor carrying plane wave and polarization contraction factors, which is irrelevant in the following calculation steps. 

Once we have the on-shell energy density, we Fourier-transform it 

\begin{equation}
    \bar{\rho}(k) = \int_0^{\infty}d\,z\, e^{ik\,z}\rho(z)
\end{equation}

\noindent to construct the \emph{modal fraction} as follows 

\begin{equation}
   f(k) =\frac{ |\bar{\rho}(k)|^2}{\int dk |\bar{\rho}(k)|^2}.
\end{equation}

The differential configurational entropy for the holographic hadron is then written as

\begin{equation}
S_{DCE}=-\int dk \,\tilde{f}(k) \log \,\tilde{f}(k)
\end{equation}

\noindent where $\tilde{f}\left(k\right)=f\left(k\right)/f\left(k\right)_\text{Max}$. The next section will discuss the DCE for some AdS/QCD models. 

\section{Bottom-up AdS/QCD test models} \label{testing}
Let us now discuss the ideas developed above in the context of bottom-up AdS/QCD models. We will consider the following static dilaton models: hard wall \cite{Boschi-Filho:2002wdj, Erlich:2005qh} softwall \cite{Karch:2006pv}, UV deformed softwall \cite{Braga:2017bml}, non-quadratic soft wall \cite{MartinContreras:2020cyg}, and UV deformed and non-quadratic softwall \cite{MartinContreras:2021bis}. 

We will consider vector mesons as testing probes: $\rho$ and $J/\psi$. Originally, the following bottom-up models were intended to describe these states. Table \ref{table:five} summarizes the data from the vector meson.

\begin{table*}[t]
\centering
\begin{tabular}{c||c||c||c||c||c}
 \hline
 \multicolumn{6}{c}{\textbf{Vector meson data}}\\
 \hline\hline
 \textbf{State} & \textbf{$I^G(J^{PC})$}&\textbf{Meson}&\textbf{Mass (MeV)}&\textbf{$\Gamma_{e^-e^+}$(KeV)} & \textbf{$f_n$ (MeV)}\\
 \hline
 \hline
 $1^3\,S_1$ & $1^+(1^{--})$ & $\rho(770)$ &$775.26\pm0.25$ & $7.04\pm0.06$& $221.21\pm0.21$\\
 \hline
  $1^3\,S_1$ & $1^+(1^{--})$ & $J/\psi$ & $3096.916\pm0.011$ & $5.547\pm0.14$& $416.16\pm5.25$\\
  \hline
  \hline
\end{tabular}
\caption{Experimental data used for $\rho$ and $J/\psi$ meson read from \cite{Workman:2022ynf}.}
\label{table:five} 
\end{table*}

\subsection{Hard Wall Model}
The hardwall model (HWM), introduced initially in the context of gluon spectrum \cite{Boschi-Filho:2002wdj} and later used to discuss the phenomenology of light vector mesons in \cite{Erlich:2005qh}, plays with the idea of inducing confinement by adding a D-brane probe, in the same sense as the square-well produces bounded states in quantum mechanics. The D-brane locus $z_{hw}$ is associated with the energy scale $\Lambda_{QCD}$ used to fix the hadron masses as follows

\begin{equation}
    z_{hw}=\frac{1}{\Lambda_{QCD}}.
\end{equation}

Since the confinement is placed by \emph{cutting} the AdS space, the static dilaton is fixed to zero in the action density \eqref{bulk-action}.

For hadrons with integer spin $S$, after  calculations, we obtain the holographic potential as 

\begin{equation}
    V_{hw}(z)=\frac{4M_5^2\,R^2 - 2\beta +\beta^2}{4z^2},\,\,\,\,0\leq z\leq z_{hw}.
\end{equation}

Let us focus on vector mesons implying $\beta=-1$ and $M_5^2\, R^2=0$. We have to adjust the cutoff $z_{hw}$ in terms of the lights unflavored vector meson, i.e., the $\rho(770)$ meson: $\Lambda_{QCD}=\frac{M_\rho}{\alpha_{1,1}}=(4.943)^{-1}$ GeV, with $M_\rho=775.26\pm0.23$ MeV \cite{Workman:2022ynf}. For scalar mesons, i.e., $\beta=-3$, the cutoff is fixed with the lightest unflavored scalar meson. 

In the case of the decay constants, the proper value of $\mathcal{K}=2\,\pi\, N_f$, with $N_f$ the number of flavors, is fixed by the large $q^2$ expansion of the 2-point function. It is expected that, at  $q^2\to \infty$, large $N_c$-QCD and AdS/QCD should have the same behavior (see \cite{Afonin:2004yb, Hong:2004sa, Afonin:2010fr, MartinContreras:2019kah}).  

\begin{center}
\begin{figure*}
   \includegraphics[width=2.3 in]{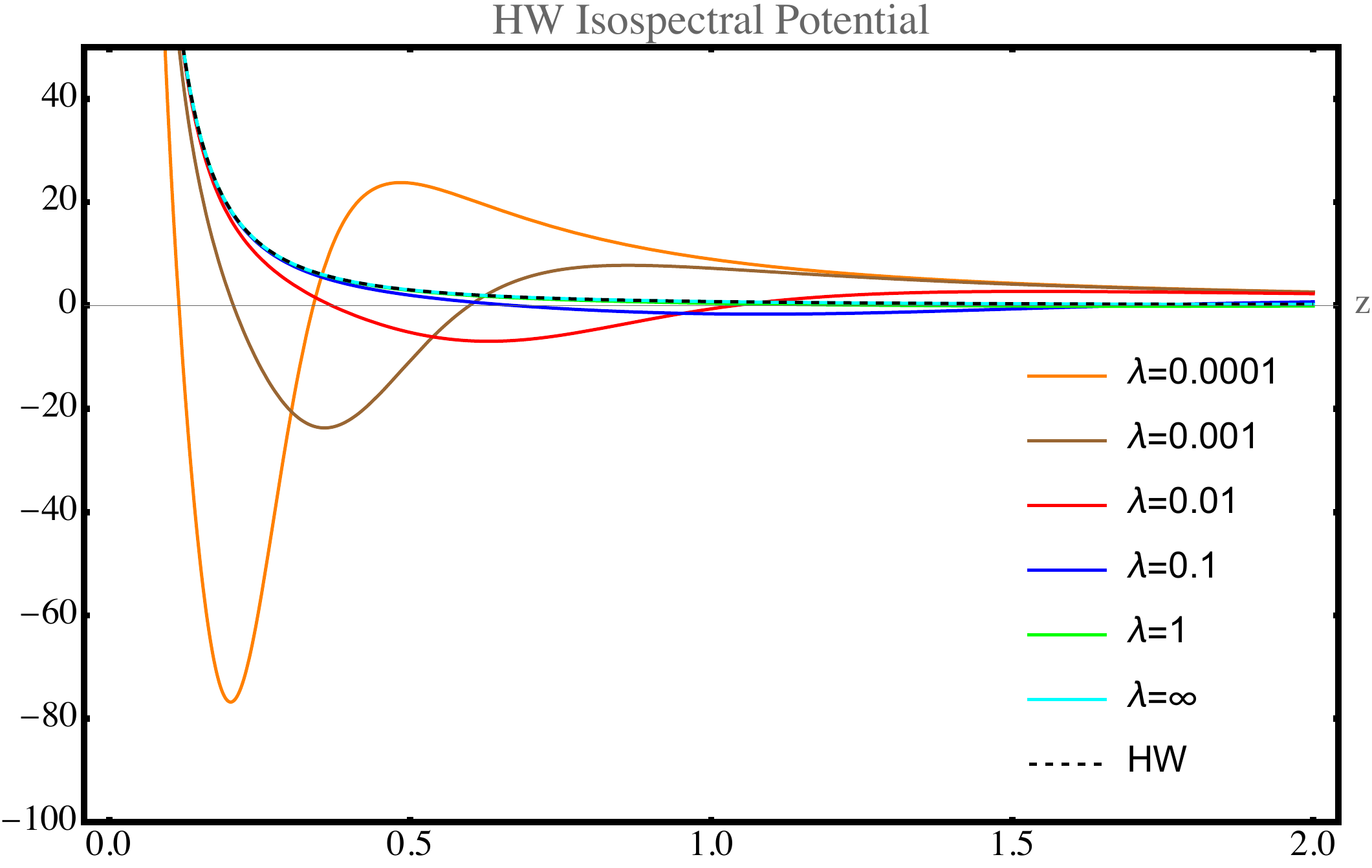} \includegraphics[width=2.3 in]{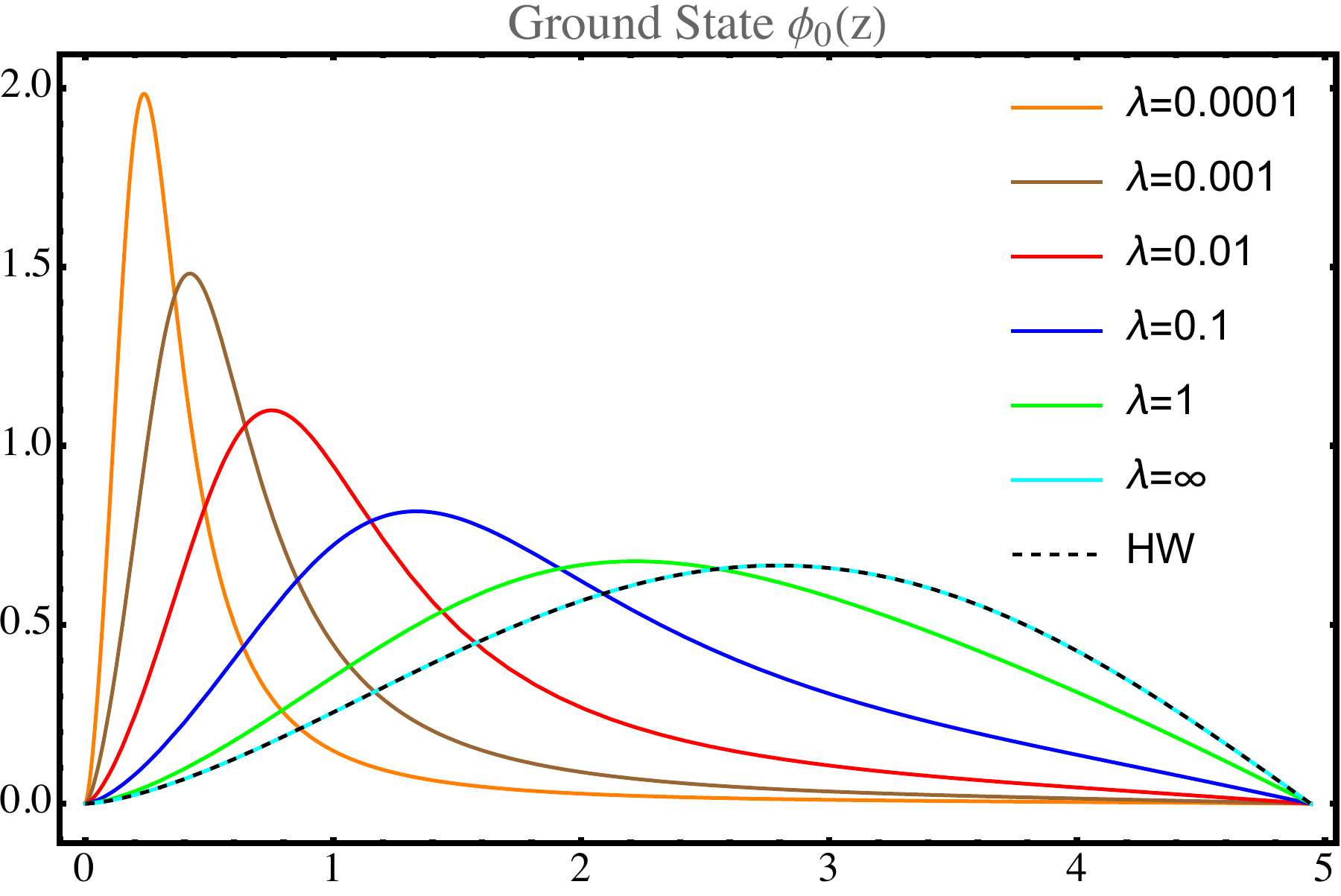} \includegraphics[width=2.3 in]{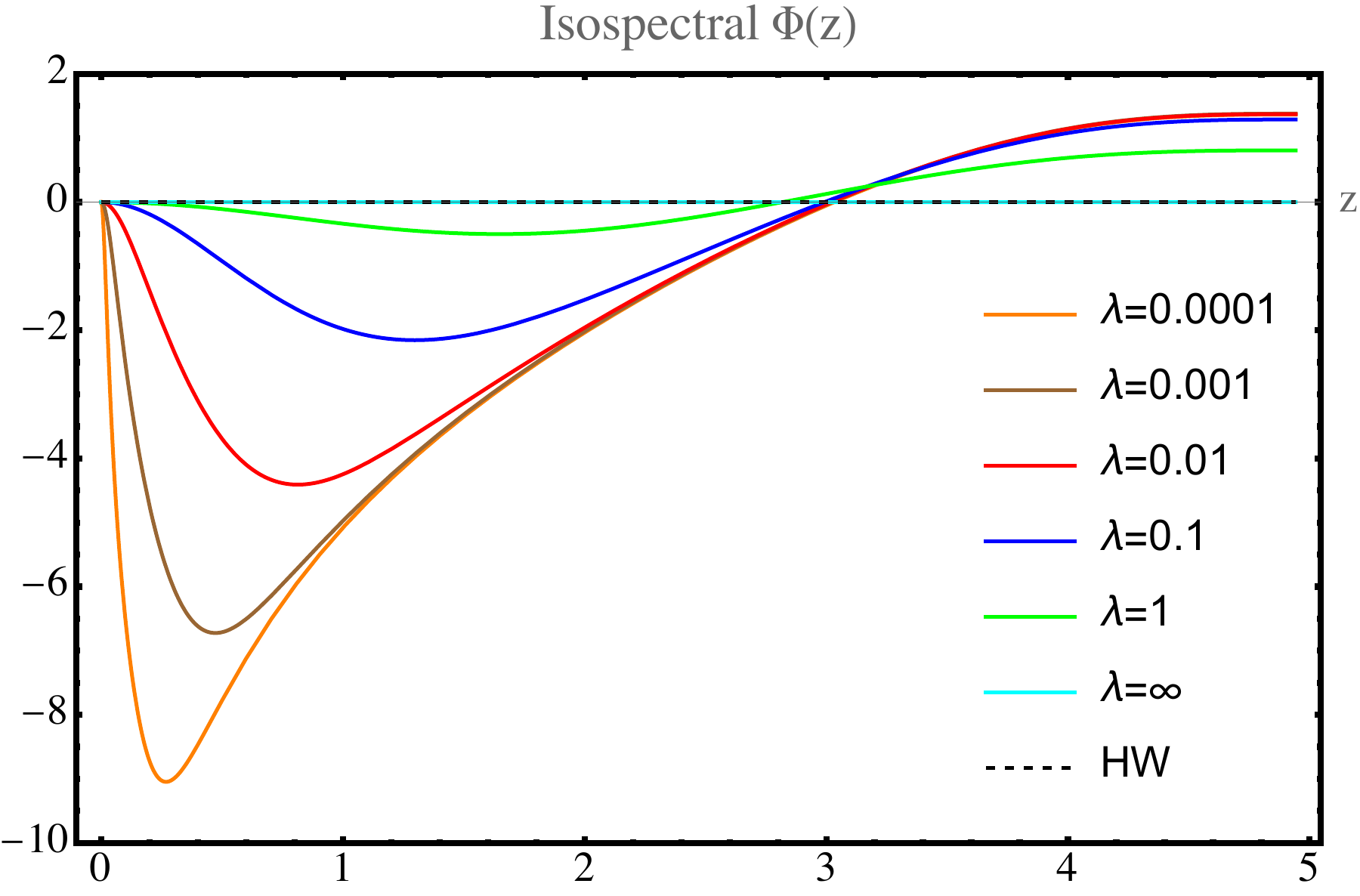}\\
\includegraphics[width=2.3 in]{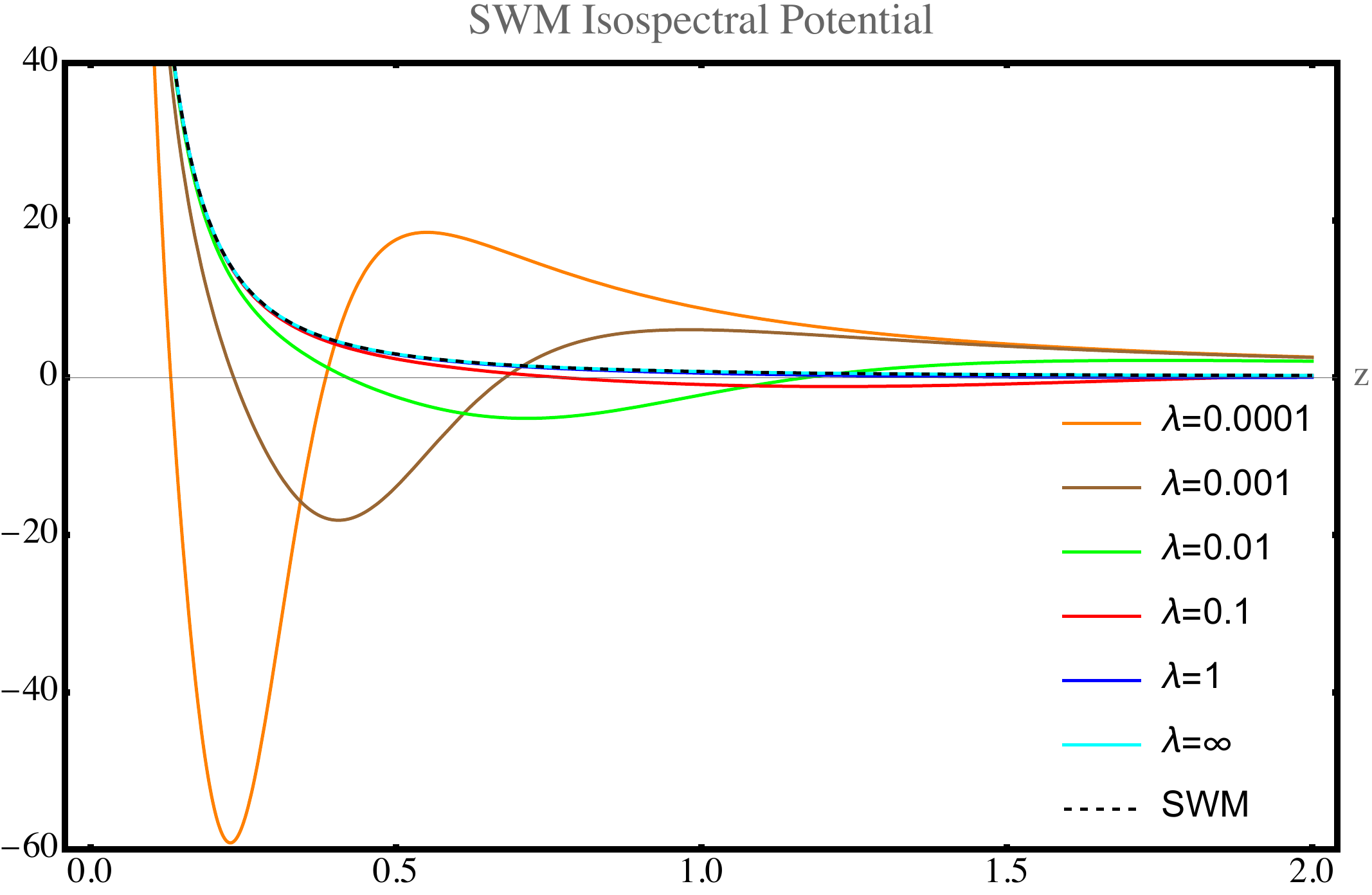} \includegraphics[width=2.3 in]{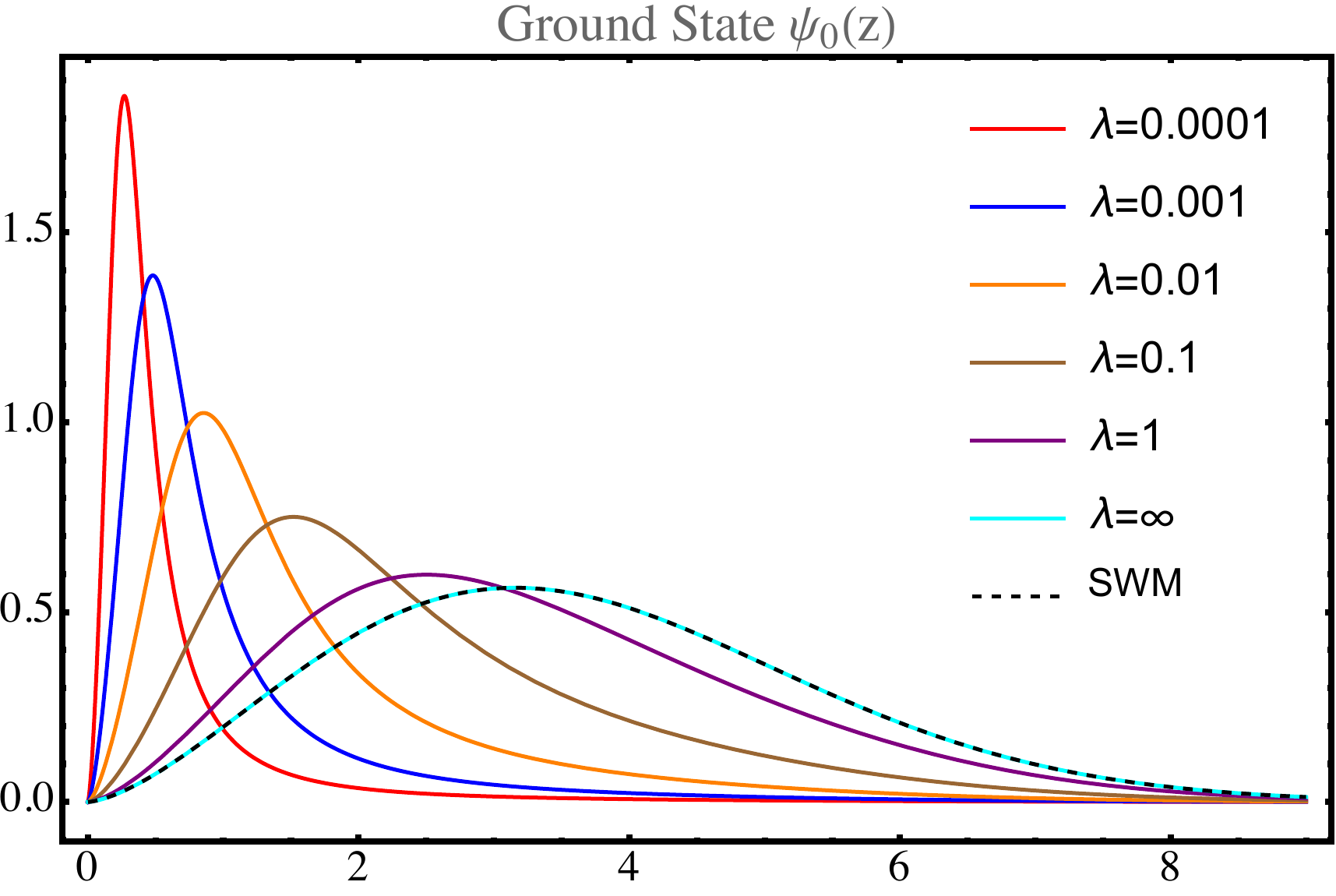} \includegraphics[width=2.3 in]{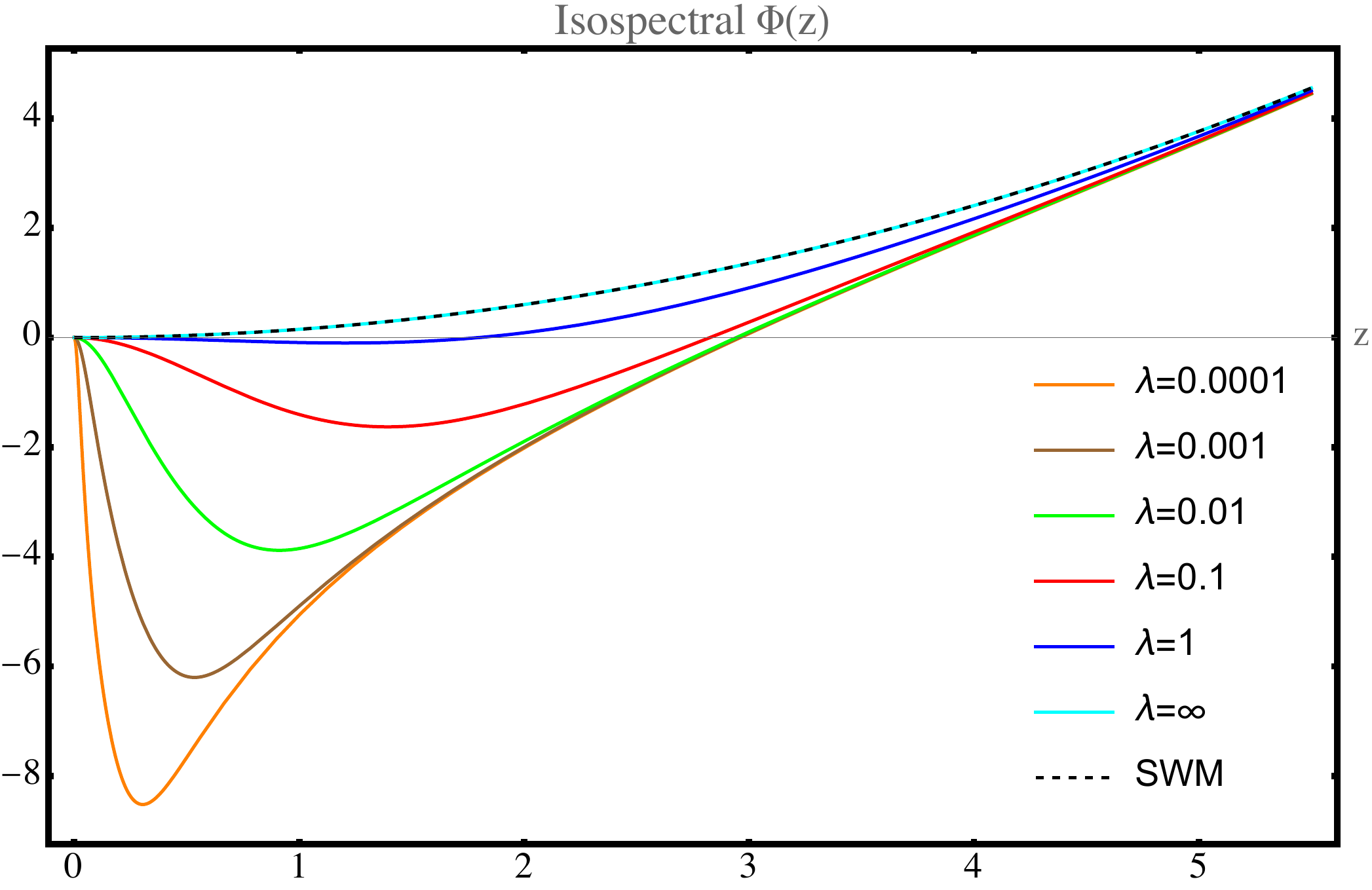}\\
\includegraphics[width=2.3 in]{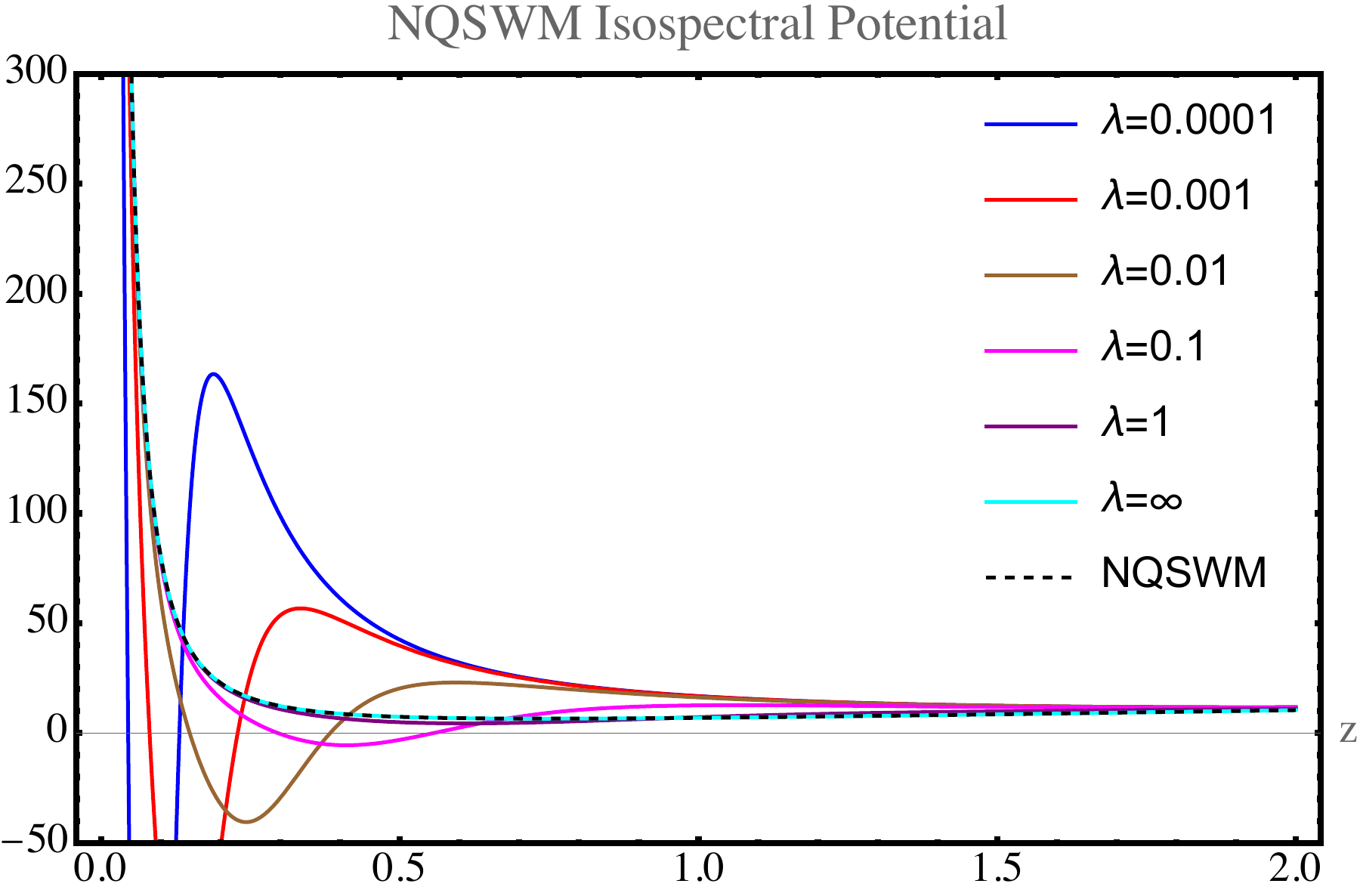} \includegraphics[width=2.3 in]{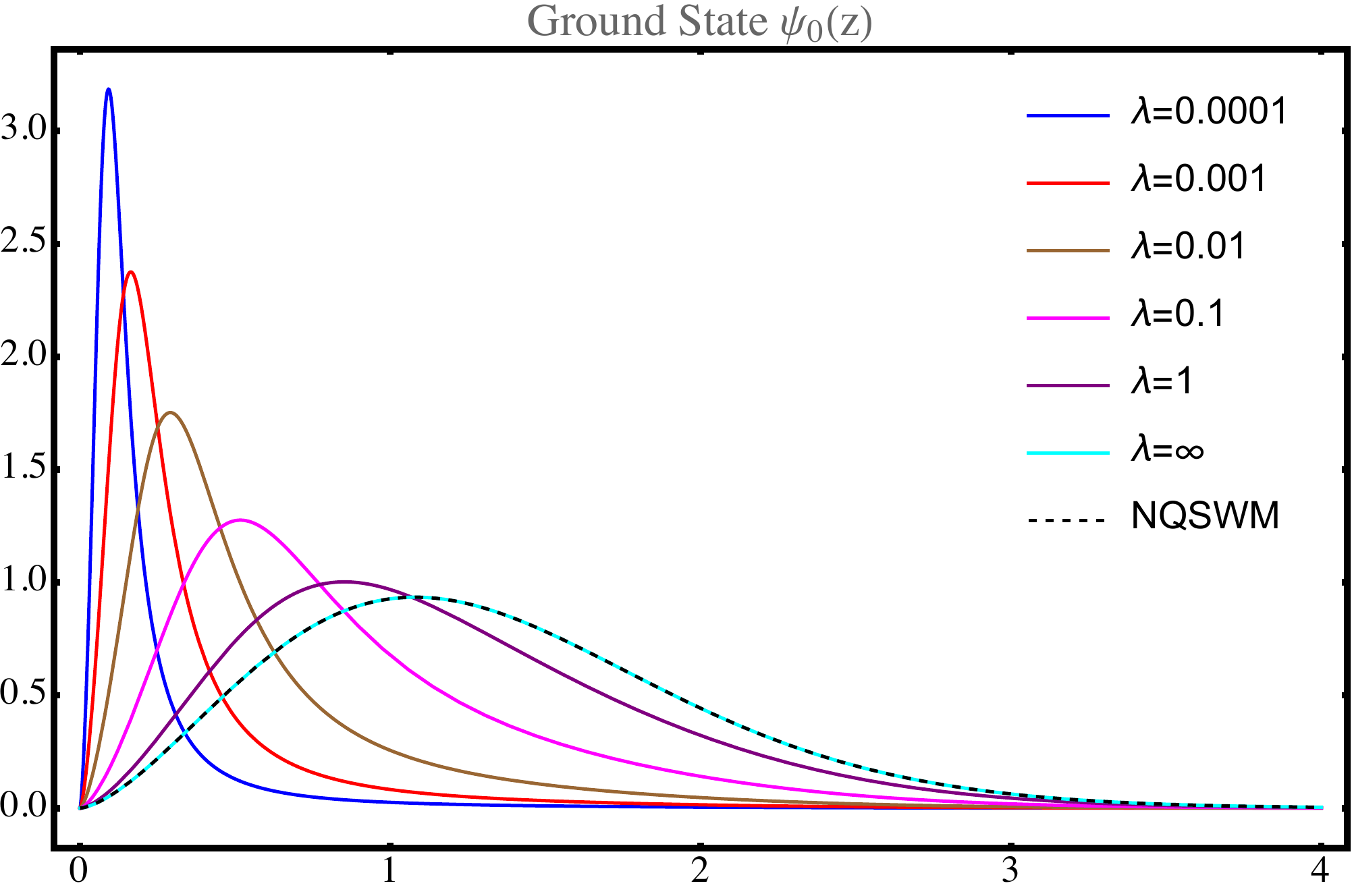} \includegraphics[width=2.3 in]{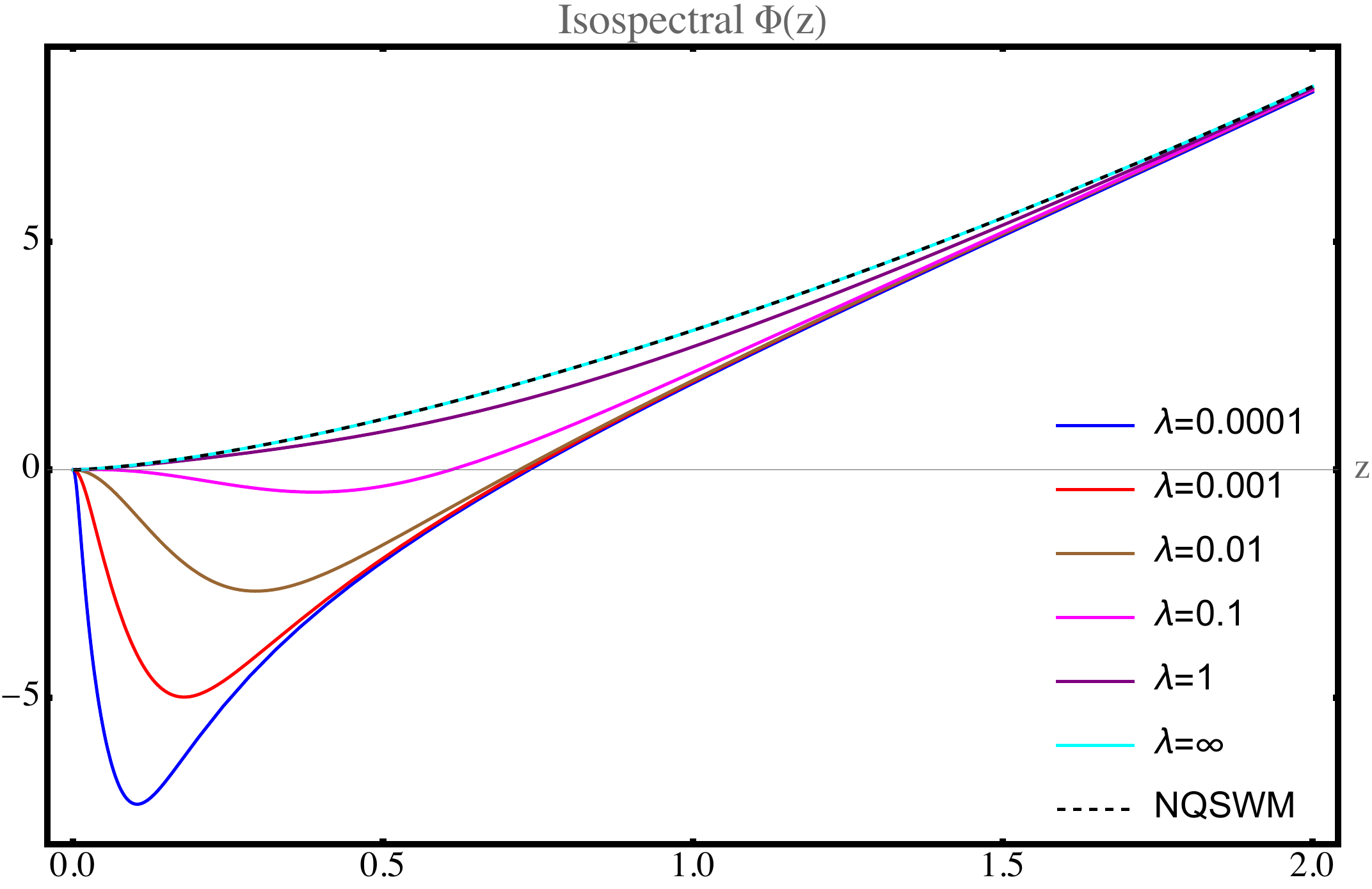}\\
\includegraphics[width=2.3 in]{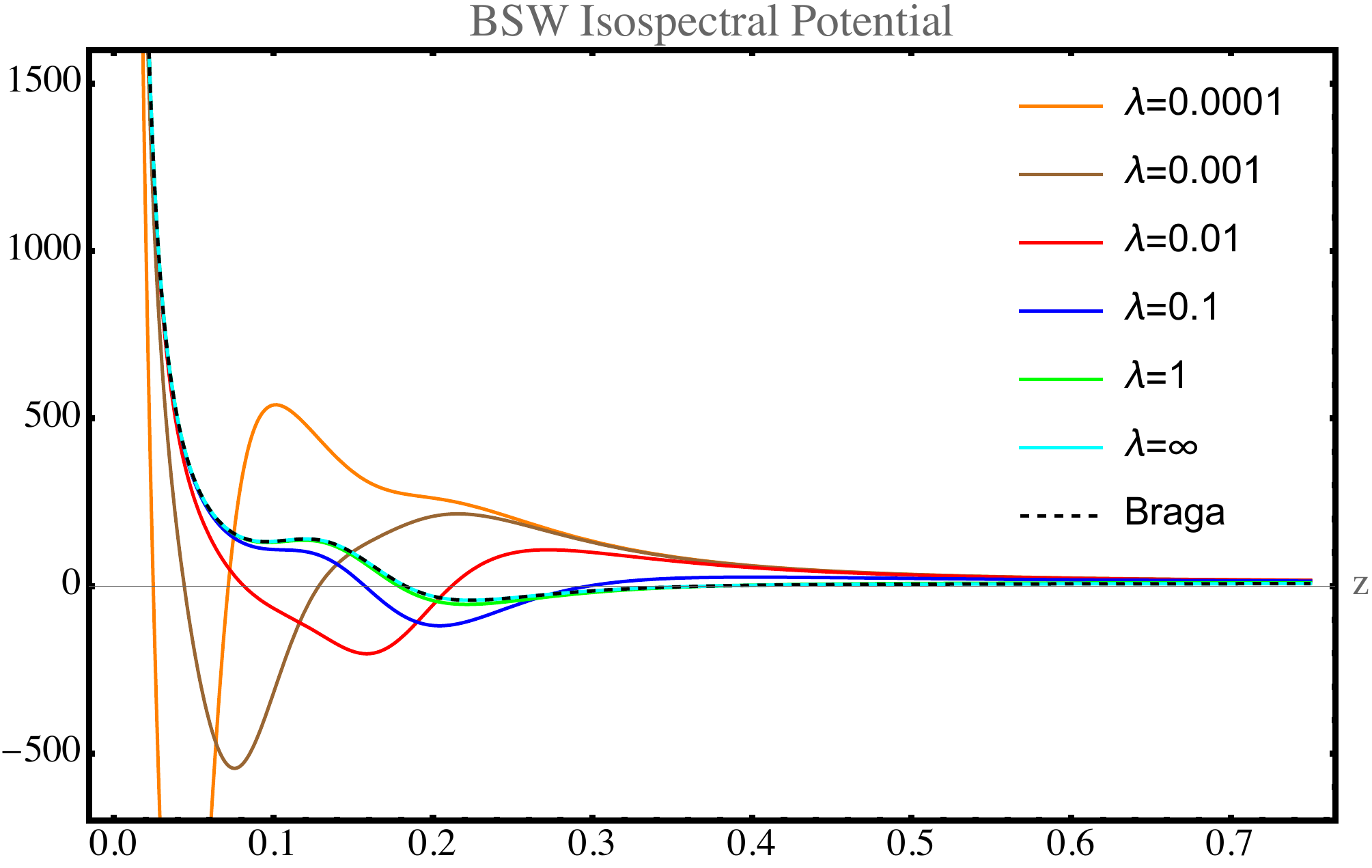} \includegraphics[width=2.3 in]{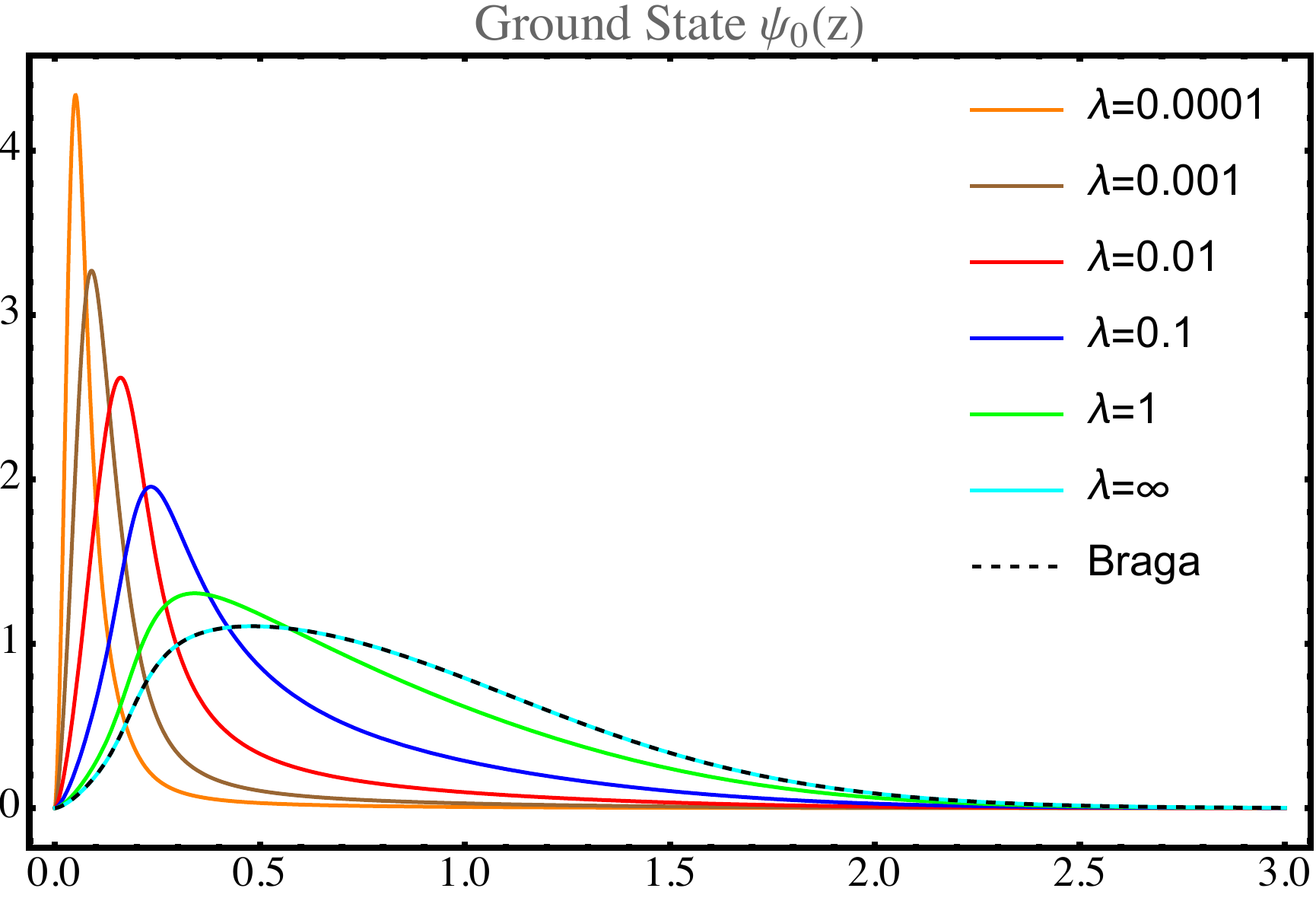} \includegraphics[width=2.3 in]{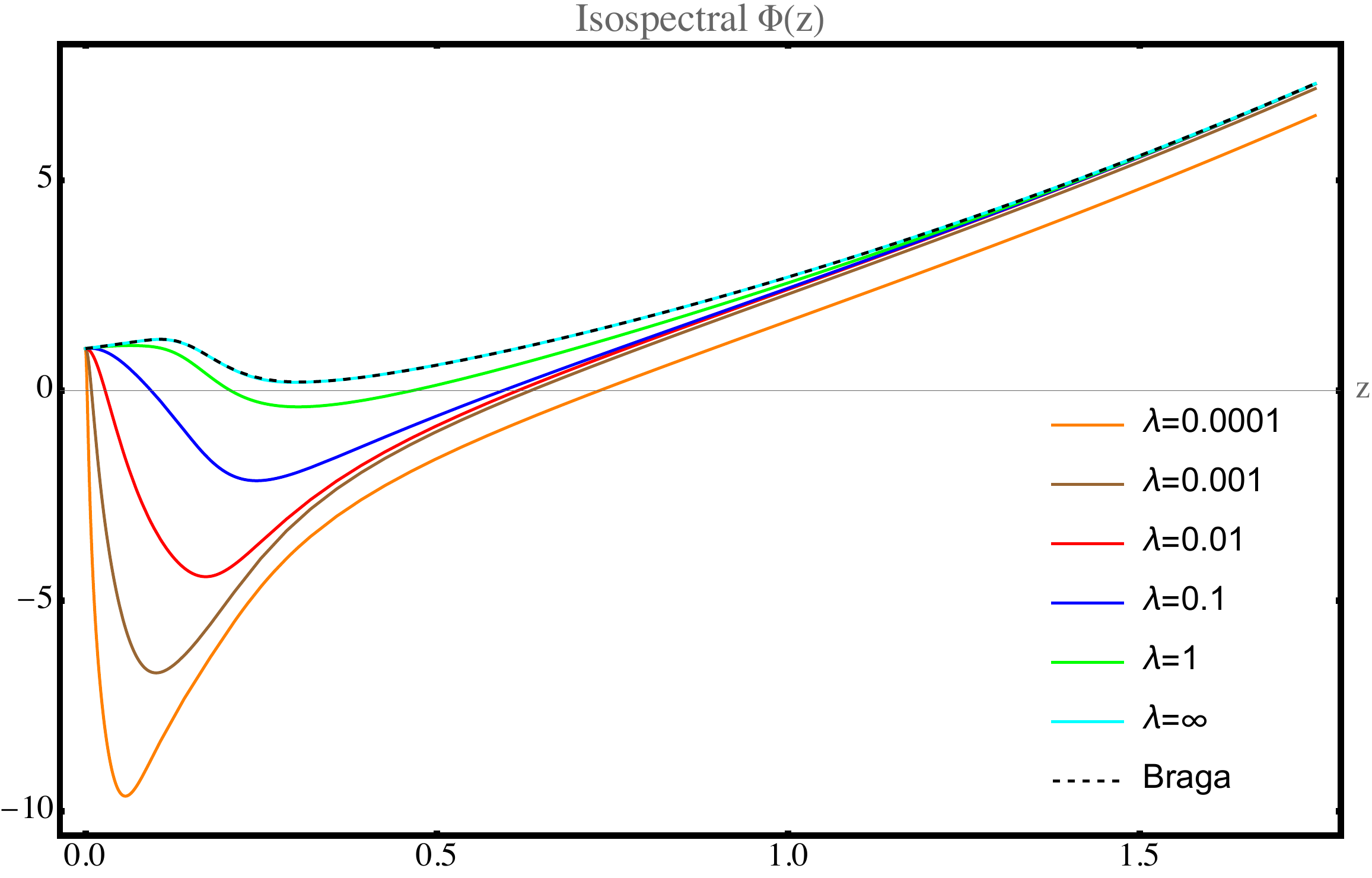}
\caption{Isospectral results for the studied holographic models discussed. In the left upper panel, we plot the family of isospectral potentials $\hat{V}_\lambda(z)$ along with the hardwall potential (dashed). In the right upper panel, we depict the Schrödinger-like ground states associated with the family of isospectral potentials. We present the isospectral dilatons $\tilde{\Phi}(z,\lambda)$ calculated from the isospectral family in the lower panel. The D-brane locus is $z_{hw}=4.949$ GeV$^{-1}$. For $\lambda=\infty$, we take $\lambda=9999$ as our "numerical infinite".}
\label{fig:one}
\end{figure*}
\end{center}

\subsection{Softwall Model}
The softwall model (SWM) was introduced in the holographic context of light vector meson spectroscopy to induce confinement by the smooth emergence of bulk-bounded states \cite{Karch:2006pv, Andreev:2006vy}. The model produces holographic hadrons by using a confining potential that, at a high $z$-value, defines the linearity of the squared mass spectrum. The confining part of the potential arises from the static dilaton profile used in the action density. In the case of the original softwall model, the dilaton is proved to be quadratic in $z$. This fact is consistent with the emergence of linear Regge trajectories at the conformal boundary. Recall that Regge trajectories are a clear signal of confinement in hadron physics. The softwall model dilaton is defined as

\begin{equation}
   \Phi_{SWM}(z)=\kappa^2\,z^2. 
\end{equation}

The lightest mass hadron in the trajectory fixes the SWM model scale $\kappa$. For the light-unflavored mesons, SWM works quite well in describing Regge trajectories. However, when you move to the heavy quarkonium realm, linearity starts to cease \cite{MartinContreras:2020cyg, MartinContreras:2021bis}. 

In general, Schrödinger-like modes are written in terms of Laguerre-associated polynomials.

The model outputs disagree with the phenomenology in the case of decay constants. In the case of vector SWM, decays are degenerated:

\begin{equation}
    f^2_n=\frac{2\,\kappa^2}{\mathcal{K}}.
\end{equation}

For scalar SWM, decays increase with excitation number. Following the same road as we did with HWM, we will focus only on vector solutions, i.e., $\beta=-1$ and $M_5^2\, R^2=0$. The energy scale is set with the lightest unflavored meson. It is customary to use $\rho(770)$ mass to fix the scale as $\kappa=0.388$ GeV. As well as the HWM, the constant $\mathcal{K}=(2\,\pi)^2$ is fixed by comparison with the large-$N_c$ two-point function at large $q^2$. 

\subsection{Braga deformed Softwall model}
Despite SWM's success in describing hadronic spectroscopy, form factors, and other phenomenology, the decay constants $f_n$ must be properly explained. \textcolor{black}{In the bottom-up models (HW and SW) framework, $f_n$ are degenerate or increasing, contrasting the experimental behavior, i.e., $f_n$ should decrease with $n$. Such a behavior is captured by the so-called \emph{Braga soft wall model}. Since decay constants depend on the eigenmode low-$z$ behavior, the dilaton field can be used to overcome this fitting problem. Such a possible dilaton is defined as}

\begin{equation}\label{nelson-dilaton}
\Phi_B(z)=\kappa^2\,z^2+M\,z\,+\tanh\left(\frac{1}{M\,z}-\frac{\kappa}{\sqrt{\Gamma}}\right).
\end{equation}

\noindent where $\kappa$ is an energy scale that controls the holographic potential high-$z$ behavior and the mass spectrum, i.e., the Regge Trajectories. The other two scales, $M$ and $\Gamma$, modify the UV behavior of the dilaton, translated into the potential also, that fixes the eigenmode low-$z$ behavior, improving the holographic decay constants: now they will be decreasing with excitation number, as it is expected from hadronic physics. However, including the UV term in the dilaton makes it lose precision in the mass spectrum. Thus, modifying the UV region implies that the dilaton field should not be quadratic. This idea is explored in \cite{MartinContreras:2021bis}.

Since the resulting isospectral potential is not analytical, it has to be solved numerically. Figure \ref{fig:one} depicts this potential with its ground state. 

This model was formulated originally for heavy quarkonia. Thus, we will fix the parameters set to fit vector charmonium. For these hadrons, we have $\kappa=1.2$ GeV, $M=2.2$ GeV, $\sqrt{\Gamma}=0.55$ GeV, $\beta=-1$, and $M_5^2\,R^2=0$. 

For decay constants,  results for vector charmonium are collected in Table \ref{table:1}.

\subsection{Non-quadratic Softwall model}
The non-quadratic softwall model (NQSWM) deformation rises from the nonlinear Regge trajectories proposed in the context of the Bethe-Salpeter equations and potential models for heavy-light mesons \cite{Chen:2018nnr, Afonin:2014nya}. This idea improves holographic spectroscopy and opens the possibility of including other non-$q\bar{q}$ hadron in a very intuitive form \cite{MartinContreras:2020cyg}. 

The model includes an extra parameter associated with the constituent mass. From WKB grounds, the high$-z$ behavior of the holographic potential controls the mass spectrum. The fact that the quadratic dilaton works fine for light unflavored hadrons implies that the SWM is chiral-symmetric by construction, yielding poor results in heavy charmonium spectroscopy. 

Then, to overtake this issue, a non-quadratic dilaton with the following form

\begin{equation}\label{Non-q-dilaton}
 \Phi_{NQ}(z)=\left(\kappa\,z\right)^{2-\alpha}
\end{equation}

\noindent allows modeling heavier hadron masses. The energy scale $\kappa$ (in GeV) usually controls the hadron mass and is associated with the strong force nature inside hadrons. In the case of the adimensional parameter $\alpha$, it controls the effect of the constituent mass in the trajectory. For light unflavored,  $\alpha=0$. For heavier mesons $\alpha\to 2/3$ (see \cite{MartinContreras:2020cyg} for a detailed discussion. This running of $\alpha$ with the constituent masses allows us to include other mesonic species as heavy-light systems or non-$q\bar{q}$ states by parametrizing $\alpha$ in terms of the quark constituent masses. Figure \ref{fig:one} depicts the holographic potential of this dilaton. 

\section{Isospectrality a la Bottom-up}\label{bottom-up iso}

The recipe exposed above allows us to explore other features of bottom-up models. By their inner nature, i.e., the dilaton choice, the results of these holographic models are expected to be phenomenologically similar. Most of the differences will rise as a result of the dilaton itself. However, the asymptotic behavior is quite different for the non-zero dilaton proposals (SW, BSW, NQSW). Recall that dilaton asymptotics determine the calculation of decay constants since the value $\Phi(z\to 0)$ acts as a normalization constant. 

\emph{For the hardwall model}, we can compute the associated dilaton field $\tilde{\Phi}_\lambda(z)$ for this isospectral family. It is interesting to note that even though in the original hardwall, the dilaton is zero, we have non-zero dilatons associated with the hardwall isospectral family. Figure \ref{fig:one} depicts the collection of isospectral dilatons. As in the case of the isospectral potentials and ground states, the isospectral dilaton tends to recover the original behavior when $\lambda\to \infty$.  

Let us apply the isospectral technology. First, using the ground state of HW, we define an isospectral function $I(z)$, given by \eqref{Darboux}.  Thus, the family of isospectral potentials is written as

\begin{equation}
 \hat{V}_{\lambda}(z)=V_{hm}(z)-2\frac{d^2}{d\,z^2}\log\left[I_{hw}(z)+\lambda\right].   
\end{equation}

We solve this potential numerically to compute the isospectral family. Figure \ref{fig:one} depicts the potentials and the associated ground states. It is interesting to notice that when we run $\lambda\to \infty$, the isospectral solutions tend to recover the original hardwall solutions. 

Regarding isospectral decay constants, the ground state decay is the only one modified by the isospectral transformation. Decay constants for excited states are not sensitive to the isospectral procedure. When we closely inspect the effect of the $I(z)$ function on the potential, the $D_z^2\,\log(I+2)$ terms affect only the low $z$ region of the holographic potential. For the regions $z\to\infty$ and $z\to0$, $V(z)$ remain unaffected. Decay constants depend strictly on the $z\to 0$ behavior of the Schrödinger mode. Thus, it is expected that isospectral and non-isospectral excited states, which were not used to build the isospectral transform \eqref{Darboux}, share the same $z\to0$ scaling behavior. Table \ref{table:1} summarizes the decay constants spectra for the different isospectral families considered. 

\emph{Let us turn our attention to the softwall model}. This is the only model with an analytical solution for the isospectral transformation, written in terms of the Gamma incomplete function $\Gamma(a,z)$. Let us prove this. To do so, start from the ground state 

\begin{equation}
    \phi_0(z)=\sqrt{2\,\kappa^4}\,z^{3/2}\,e^{-\frac{1}{2}\,\kappa^2\,z^2}.
\end{equation}

Then, the isospectral transformation takes the following form 

\begin{equation}
    I_{SW}(z)=2\,\kappa^4\,\int_0^z{dx\,z^3\,e^{-\kappa^2\,x^2}}=1-\Gamma\left(2,\kappa^2\,z^2\right),
\end{equation}

\noindent that is written using the incomplete Gamma function. We can perform this calculation to prove that the expression above is independent of the hadronic spin. The monoparametric isospectral potential is now written as

\begin{multline}
    \hat{V}_{SW}(z)=V_{SW}(z)\\
    -2\,\frac{d^2}{d\,z^2}\,\log \left[1-\Gamma\left(2,\kappa^2\,z^2\right)+
    \lambda\right].
\end{multline}

The family of isospectral potentials for the vector softwall model is depicted in Figure \ref{fig:one}, along with the dilaton and isospectral ground states. 

Compared to HW, the SW model has the same isospectral behavior (nor the same holographic phenomenology). For $z\to0$ and $z\to\infty$, the isospectral potential behaves as its SW counterpart. However, in the intermediate region, the isospectral potential behaves differently. 
As in the case of the HW model, $I_{SW}(z)$ strongly modifies the decay constant of the ground state, leaving the excited ones unchanged. For vector mesons, isospectral transformations break the degeneracy only for the ground state. Excited states keep being degenerated. Isospectral decay constants are summarized in Table \ref{table:1}.

In light-front holography with quadratic dilaton, Brodsky \emph{et al.}  have discussed using SUSY algebra to accommodate mesons, baryons, and tetraquarks in a supermultiplet. See \cite{Brodsky:2016yod, Brodsky:2016rvj, Nielsen:2018uyn, Nielsen:2018ytt} and references therein.

\begin{table*}[t]
\centering
\begin{tabular}{c||c|c|c|c||c|c|c|c||c|c|c|c||c|c|c|c}
 \hline
 \multicolumn{17}{c}{\textbf{Isospectral decay constants $f_n$ (GeV)}}\\
 \hline\hline
 \multicolumn{1}{c||}{$\lambda$} & \multicolumn{4}{c||}{Hardwall Model} & \multicolumn{4}{c||}{Softwall Model}&\multicolumn{4}{c||}{Braga SW Model} & \multicolumn{4}{c}{Non-quadratic SW Model} \\ 
 \cline{2-2} \cline{3-3} \cline{4-4} \cline{5-5}\cline{6-6} \cline{7-7} \cline{8-8} \cline{9-9}\cline{10-10} \cline{11-11} \cline{12-12} \cline{13-13}\cline{14-14} \cline{15-15} \cline{16-16} \cline{17-17}
 &$f_0$ & $f_1$ & $f_2$ & $f_3$ & $f_0$ & $f_1$ & $f_2$ & $f_3$&$f_0$ & $f_1$ & $f_2$ & $f_3$ & $f_0$ & $f_1$ & $f_2$ & $f_3$\\
 \hline\hline
 $10^{-4}$  & 11.346 & 0.151 & 0.182 & 0.208 & 8.742  & 0.087 & 0.087 & 0.087 & 42.685& 0.226 & 0.187 & 0.163 & 19.046& 0.177 & 0.175 & 0.171\\
 $10^{-3}$  & 3.538  & 0.150 & 0.180 & 0.206 & 2.744  & 0.087 & 0.087 & 0.087 & 12.744 & 0.249 & 0.196 & 0.168 & 5.802 & 0.179 & 0.174 & 0.169\\
 $10^{-2}$  & 1.134  & 0.152 & 0.182 & 0.208 & 0.876  & 0.087 & 0.087 & 0.087 & 4.001 & 0.255 & 0.198 & 0.169 & 1.889 & 0.184 & 0.178 & 0.173\\
 $10^{-1}$  & 0.374  & 0.152 & 0.182 & 0.208 & 0.289  & 0.087 & 0.087 & 0.087 & 1.322 & 0.255 & 0.198 & 0.169 & 0.625 & 0.185 & 0.179 & 0.174\\
 1   & 0.159  & 0.152 & 0.182 & 0.208 & 0.123  & 0.087 & 0.087 & 0.087 & 0.564 & 0.255 & 0.198 & 0.169 & 0.266 & 0.185 & 0.179 & 0.174\\
 9999  & 0.113  & 0.152 & 0.182 & 0.208 & 0.087  & 0.087 & 0.087 & 0.087 & 0.399 & 0.255 & 0.198 & 0.169 & 0.188 & 0.185 & 0.179 & 0.174\\
 Non. Iso. & 0.113  & 0.152 & 0.182 & 0.208 & 0.087  & 0.087 & 0.087 & 0.087 & 0.399 & 0.255 & 0.198 & 0.169 & 0.188 & 0.185 & 0.179 & 0.174\\
 \hline \hline
\end{tabular}
\caption{Decay constants calculated using isospectral models. For hardwall, the pion mass has fixed the D-brane locus. For the softwall model, we have fixed $\kappa=0.388$ GeV for the light unflavored meson family using the pion mass. In the case of Braga SW, we used parameters $\kappa_B=1.2$ GeV, $M_B=2.2$ GeV, $\Gamma_B=0.55^2$ GeV to fit charmonium spectra \cite{Braga:2016wzx}. For the non-quadratic softwall, we choose $\kappa_{nl}=2.150$ GeV and $\alpha_{nl}=0.54$ to fit the charmonium spectrum \cite{MartinContreras:2020cyg}.}
\label{table:1}
\end{table*}

\emph{Let us consider the Braga deformed SWM}. The isospectral potential is written from eqn. \eqref{isopotential} numerically. In figure \ref{fig:one}, we plot the isospectral family of potentials $\hat{V}_\lambda(z)$ related to this holographic potential. As in the previous cases, the isospectral transformation modifies the low-$z$ potential behavior. For the excited states, the eigenfunctions in the family have the same slope in the limits $z\to 0$ and $z\to \infty$. The decay constants spectrum supports this observation (see Table \ref{table:1}), where only the ground state decay is modified.

\begin{table}[t]
\centering
\begin{tabular}{c||c||c}
 \hline
 \hline
\multicolumn{3}{c}{\textbf{Vector meson decay constants}}\\
 \hline\hline
 \textbf{Model [Meson]}&\textbf{$f_n\,(\%\,\text{R. E.})$ in Mev} &\textbf{$\lambda$}\\
 \hline \hline 
 SWM [$\rho(770)$] & $226.29(2.3)$ & $0.175$\\
 \hline
 HWM [$\rho(770)$] & $222.04(0.37)$ & $0.35$\\
 \hline
 \hline
BSWM [$J/\psi$] & $418.99(0.68)$ & $9.5$\\
\hline
NQSWM [$J/\psi$] & $415.33(0.20)$& $0.26$\\
\hline \hline
\end{tabular}
\caption{Summary of the isospectral results for the ground states of $\rho$ and $J/\psi$ mesons with the specific values of $\lambda$ for each model. In parenthesis, we write the relative error with the experimental data in Table \ref{table:five}.}
\label{table:seven} 
\end{table}

A similar situation occurs with the dilaton field. Despite the original idea of using this deformation to improve the SWM decay constants, the isospectrality modifies the ground state only, leaving the decay constants with higher $n$ unperturbed. The most significant modifications are observed in the low $z$ region, where the dilaton has a pronounced deformation for $\lambda \to 0$, as in other AdS/QCD models. 

\emph{For the non-quadratic SW}, the story is not that different. The isospectral family of potential $\hat{V}_{\lambda}(z)$ is defined numerically using eqn. \eqref{isopotential}. In Figure \ref{fig:one}, we plot the isospectral family of potentials, ground states, and isospectral dilatons for this deformed model. The isospectral potential only modifies the low $z$ behavior of the dilaton fields. In the limits $z\to0$ and $z\to \infty$, isospectral dilatons behave as the non-isospectral counterpart: vanishing for $z\to 0$ and $z^{2-\alpha}$ for $z\to \infty$. 

For the Schrödinger modes, decay constants are modified for the isospectral ground states. As in the previous cases, this isospectral transformation did not affect the excited modes decay constants, supporting that the isospectral transformation does not affect the low-$z$ behavior of excited states. Table \ref{table:1} presents the summary of isospectral decay constants.  

At this point of the discussion, it is worth mentioning that the observed behavior of the isospectral decay constants is customary for the AdS/QCD models considered here, and it can be generalized to any other SW-like model, i.e., another dilaton-based model. In fact, in the holographic potential, dilaton derivatives control the high-$z$ asymptotics of the holographic potential. The AdS Poincare patch warp factor, unaffected by isospectral transformation, dominates the low-$z$ region. 

Once we have been convinced how isospectrality modifies the decay constant for the ground state of the meson, we fit the experimental decay constants summarized in Table \ref{table:five}. The results are exposed in Table \ref{table:seven}.  

As a general conclusion, all of the holographic quantities calculated with the isospectral states tend to their non-isospectral counterpart when $\lambda\to \infty$.

\begin{figure}
  \includegraphics[width=3.4 in]{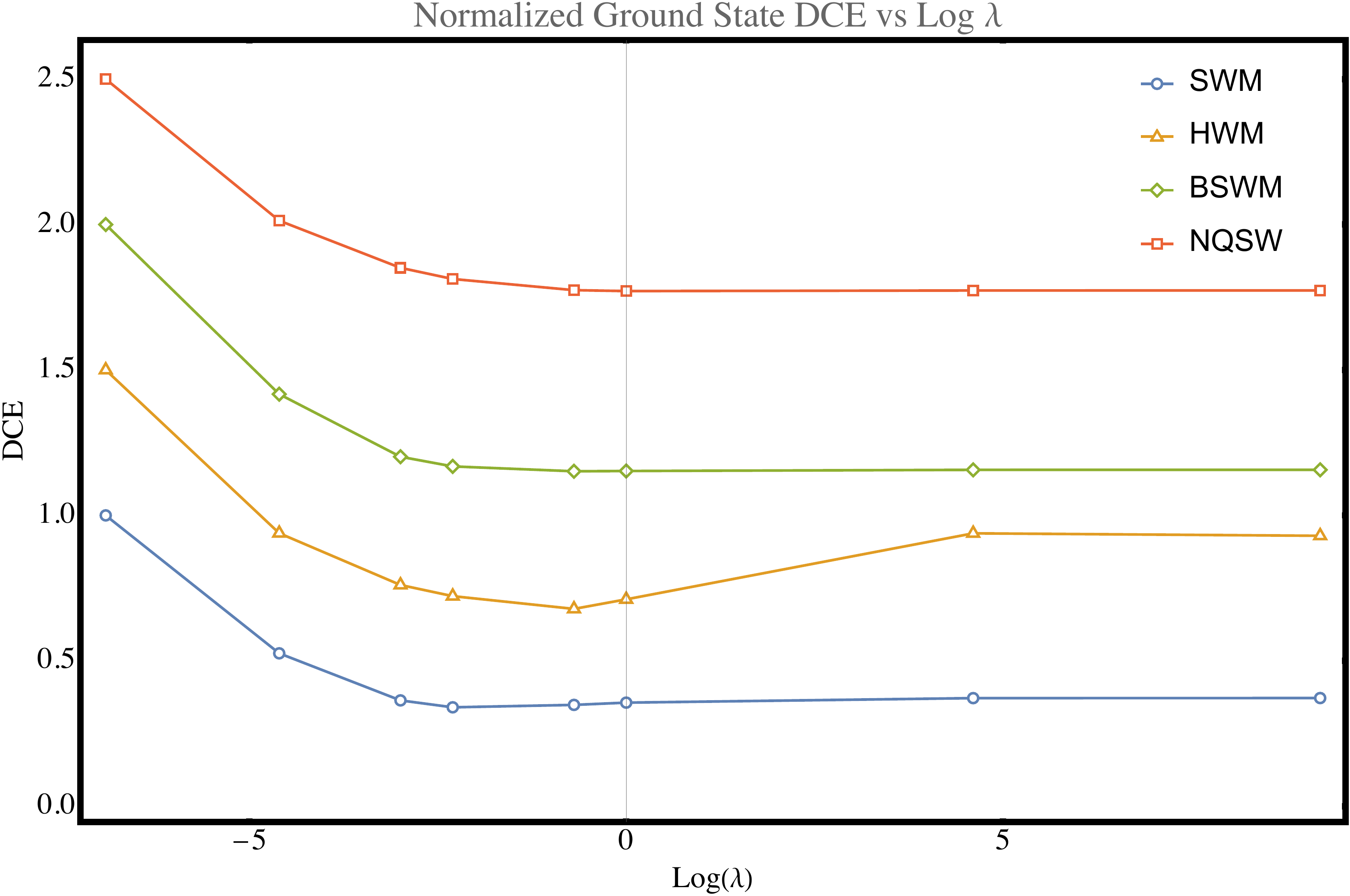}
\caption{DCE for the ground state for each AdS/QCD model considered, as a function of the $\lambda$ considered.}
\label{fig:five}
\end{figure}

\begin{figure*}
  \includegraphics[width=2.6 in]{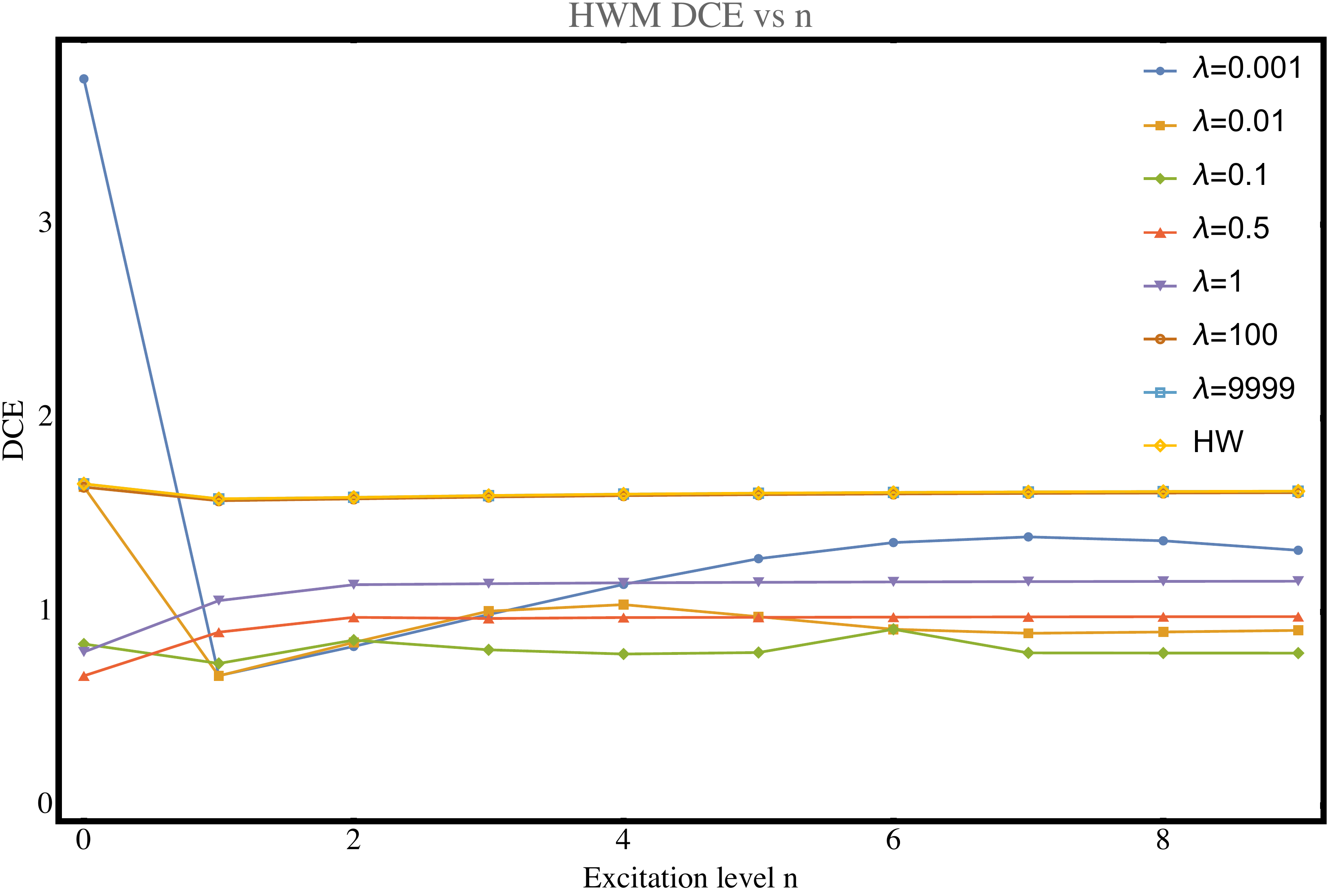}
  \includegraphics[width=2.6 in]{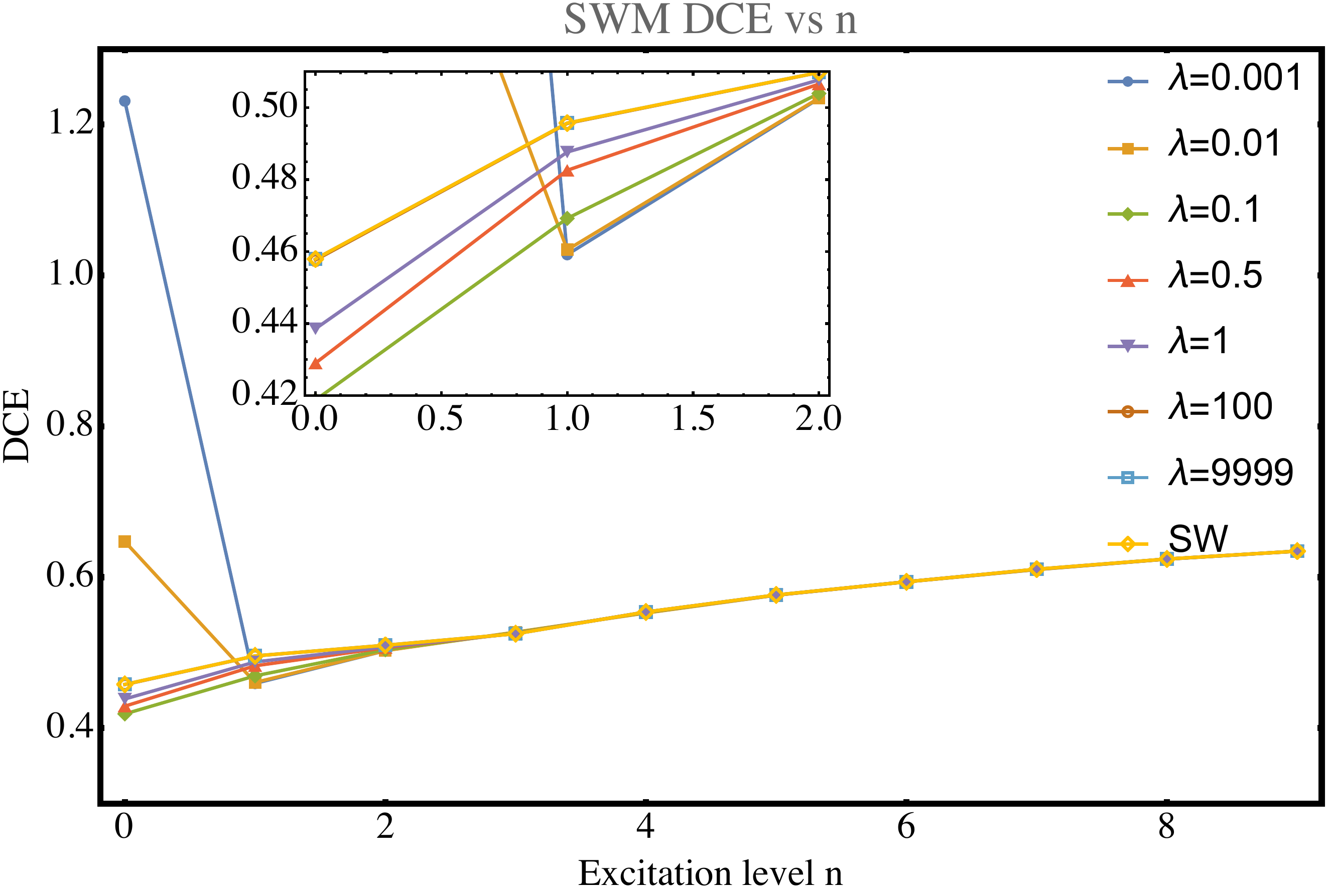}
  \includegraphics[width=2.6 in]{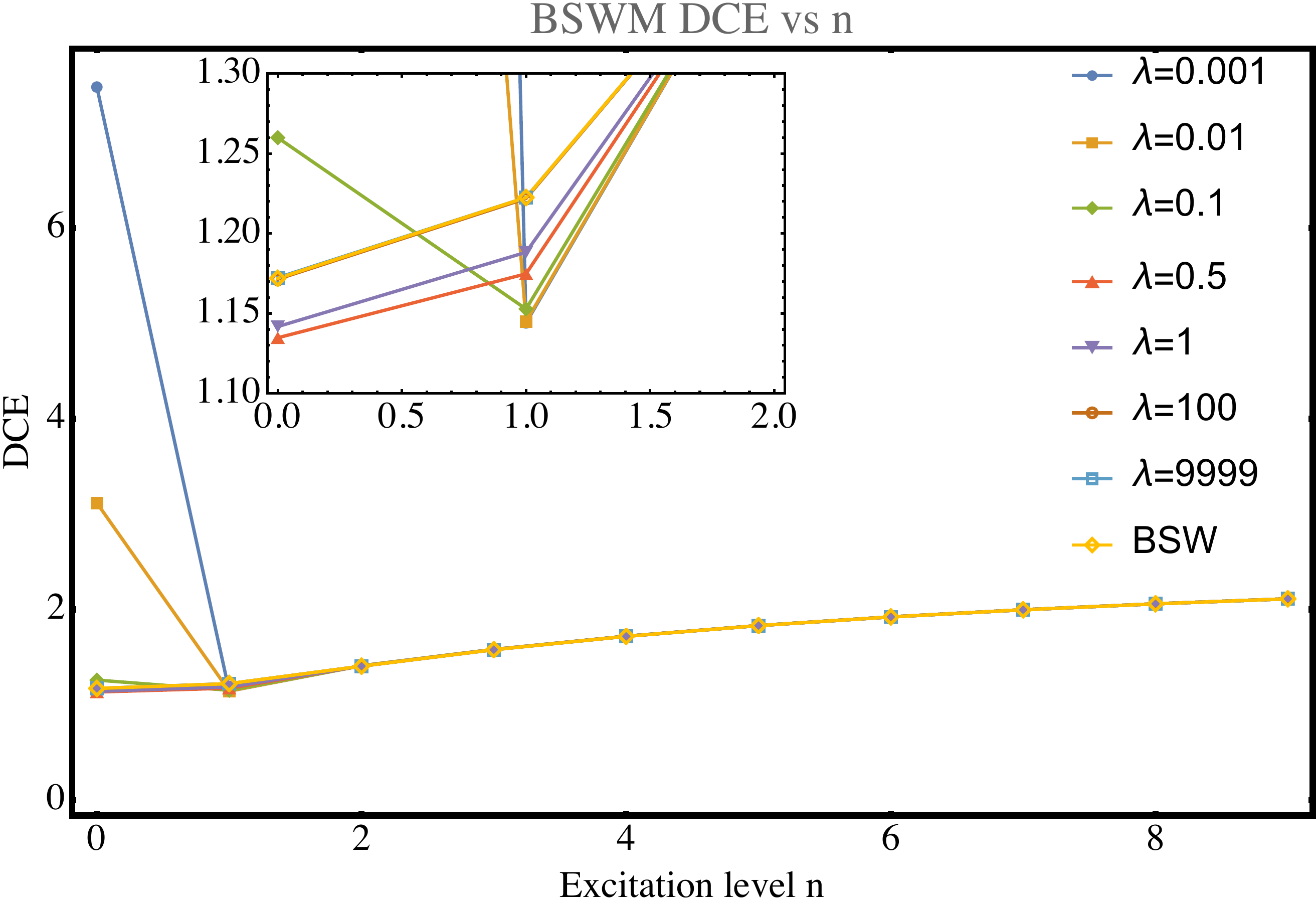}
  \includegraphics[width=2.6 in]{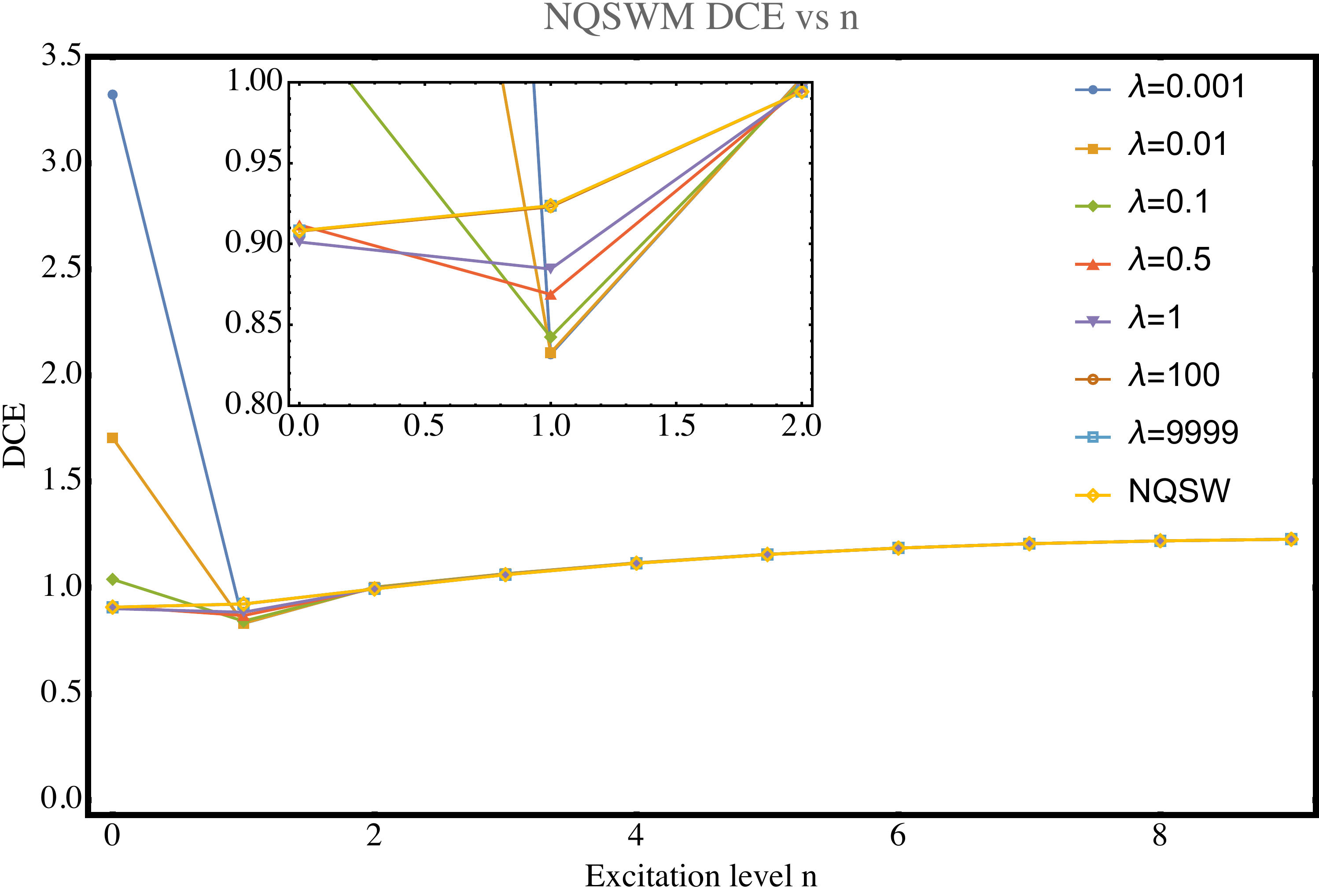}

\caption{DCE for the mass spectra in each isospectral family. Isospectrality breaks the stability in these models. In the hardwall model (HWM) case, for $\lambda\to0$ ($\lambda=0.001,\,0.01,\,0.1$), ground states are more unstable than the excited ones, implying that these isospectral families lead to inconsistent holographic results compared to the boundary data supported by experiments. When $\lambda\to1$ ($\lambda=0.5,\,1.0$), ground states become more stable than the excited modes. For dilaton-based models, the situation is similar. For the softwall model (SWM), $\lambda=0.001,\,0.01$ leads to instabilities as those observed for HWM. Braga (BSWM)  model has the same behavior as HWM, i.e., $\lambda=0.001, \, 0.01,\,0.1$ leads to instability.  The non-quadratic soft wall (NQSWM) is unstable for $0<\lambda<1$.}
\label{fig:four}
\end{figure*}

\section{Differential Configurational Entropy and  Isospectrality}\label{Isos and DCE} 
Differential configurational entropy becomes handy to analyze the connection between isospectrality and stability for these models at zero temperature. In the last section, we saw how isospectrality modifies the dilaton behavior. For effective model engineering, isospectrality can open up new possibilities to encode (or decode) holographic information. 

We shall consider the DCE behavior for ground states and excited states. We expect the ground states to have the smallest DCE. These analyses will lead us to define a criterium of \emph{which isospectral family will be suitable to describe stability in the mesonic tower of states}. A summary of our results is given in Figure \ref{fig:four}. We use non-isospectral DCE as control data since we expect isospectral DCE to meet non-isospectral models at $\lambda \to \infty$. 

For the first test, the behavior of all four models is consistently the same. The $\lambda\to\infty$ ground mode does not reach the lowest DCE value. We observed that when $\lambda\to0$, DCE in the ground states tends to increase. 

From purely entropic grounds, it was expected that, among the isospectral family, the ground state with the lowest CE would be preferred. This is not the strict rule in all the studied cases. 

In the case of the second test, when $\lambda\to 0$, in addition to the ground state, the excited states increase their DCE with the excitation number $n$ (see Fig. \ref{fig:four}). It is worth noticing the difference between HW and SW-like models. DCE for isospectral HW demonstrates that the most favorable solution is for $\lambda=1$. DCE for $\lambda\to 0$ shows that the model is unstable. When the dilaton field starts to play, DCE has a particular and fascinating behavior. Isospectrality seems to affect the ground state over the excited ones since these states have very close DCE despite the isospectral family since the bulk mode and mass are similar. 

Isospectrality accommodates the fit for the ground state decay constant while increasing the DCE with the excitation level. This stability criterion for  AdS/QCD models simultaneously describes experimental data (decay constant) and meets that excited states are more unstable (DCE). The former is an objective criterion since the ground state decay constant comes from experiments,  as happens for the light meson $\rho$ and the heavy meson $J/\Psi$. The latter is more subjective since no experimental data on DCE for vector mesons exists. However, the connection between DCE and stability is well-established in the literature.  Increasing DCE means losing stability. This observation directly relates to high-excited states being harder to observe than ground states in Nature. 

\textcolor{black}{This observation can be supported by analyzing the electromagnetic decay width $\Gamma$ of vector mesons going into leptons. An expression for the decay width and DCE from the Segré formula can be deduced. Following proposals as Ref. \cite{daRocha:2023nsb}, it is possible to find the hadronic spectrum as a function of the configurational entropy. For hadronic states, we expect $M_n(DCE)\sim DCE^\tau\sim n^\gamma$, where $\tau,\,\gamma>0$. Note that $\gamma$ is a parameter coming from fitting the hadronic mass in terms of DCE. Since DCE is an increasing function of $n$ for hadrons. From the Fermi-Segre formula (see, for example, Ref. \cite{Quigg:1977wn}), we find the connection between the hadronic biding energy $E_n$ and the wave function in the origin $\left|\psi(0)\right|$ at the conformal boundary, as follows:}

\begin{equation}
\Gamma \sim\left|\psi(0)\right|^2\sim E_n^{1/2}\,\frac{d\,E_n}{dn}
\end{equation}

\textcolor{black}{In parallel, we can define the hadron mass, in terms of the binding energy and the constituent mass, as $M_n=M_1+M_2+E_n$. Thus, the Segre formula can be written approximately as} 

\begin{equation}
\Gamma \sim\left|\psi(0)\right|^2\sim M_n^{1/2}\,\frac{d\,M_n}{dn}\sim \gamma \,n^{\frac{3}{2}\gamma-1}
\end{equation}

\textcolor{black}{This imposes a phenomenological bound for the constant $\gamma$, i.e., $0\leq \gamma <2/3$ for models having stable hadrons. The proper value of $\gamma$ is defined by the relation between hadron mass and DCE and also by the linearity of the Regge trajectory. The most general trajectory to prove this affirmation is $M_n^2=a(n+b)^\nu$. In such  a case, we have $M_n^{1/2}\sim a^{1/2}\,n^{\nu/4}$ and $\frac{d\,M_n}{d\,n}\sim a\,\frac{\nu}{2}\,n^{\nu/2-1}$, and therefore, for the decay width we conclude}

\begin{equation}
  \Gamma \sim \left|\psi(0)\right|^2\sim \nu\, a^{3/4}\,n^{\frac{3}{4}\nu-1}.
\end{equation}

\textcolor{black}{Thus, we can find a connection between the DCE exponent $\gamma$ and the linearity $\nu$ by comparing the expressions for the decay width: $\nu=2\,\gamma$. For linear Regge trajectories, i.e., $\nu=1$ (and $\gamma=1/2$), we obtain the following result for the decay width}

\begin{equation*}
    \Gamma\sim n^{-1/4},
\end{equation*}

\noindent \textcolor{black}{which is consistent with the Van Royen-Weisskofpt formula \cite{Lucha:1991vn}. Thus, models with increasing DCE with $n$ and hadronic mass $M_n$ describe stable hadrons. }

\section{Conclusions}\label{conclusions}
This work discusses how isospectrality, differential configurational entropy, and AdS/QCD models are connected. Ref. \cite{Cooper:1994eh} summarizes the machinery used to build supersymmetric partners from a given quantum mechanics potential. However, this formulation raises whether you have two potentials belonging to an isospectral family. Which one should be preferred to the other when modeling a given phenomenology? Suppose that one has a Coulomb potential. In this particular Coulomb example, the potential structure comes from well and deep known electromagnetic phenomenology. 

In holographic QCD, effective models have more phenomenological "freedom", allowing the existence of many proposals. At this point, isospectrality adds more ingredients to the story. Any isospectral potential, with a nonspecified $\lambda$, can generate any AdS/QCD model, according to the master equation \eqref{dialton-eng}. In this context, the above question found a new soil to grow: Among the possible AdS/QCD models we can develop, which ones can adequately describe hadronic spectroscopy? The answer comes from other observables associated with the hadronic stability, measured in the decay constant. We find that isospectrality maps a tower of states that share equal mass spectra with different decay constants for ground states. In this sense, the isospectrality provides a natural form to select a potential in the equivalence class by matching with the experimental decay constants. 

The discussion on stability in a class of isospectral holographic potentials is enriched by analyzing another observable: differential configurational entropy. In practice, DCE measures the hadronic state stability. The bigger the DCE, the more unstable the associated state will be. For this reason, a suitable AdS/QCD model for hadronic spectra should provide a monotonically increasing DCE with hadronic radial excitation. On the other hand, we also desire the AdS/QCD model to precisely fit the ground state decay constant, which can be done by adjusting the isospectral parameter. A suitable AdS/QCD model for mesons is expected to be consistent with both observables. A direct comparison with the results of Table \ref{table:seven} leads to the conclusion that the hardwall, the softwall, and the Braga deformed SW models are simultaneously consistent with the ground state decay constant and the DCE hierarchy since the values of $\lambda$, providing the fitting for the decay constant, rely on the acceptable regions of DCE. However, the non-quadratic softwall model fails to satisfy both requirements since the fit of $J/\psi$ decay constant leads to a ground state more unstable than the first excited state.

The existence of an equivalence class of isospectral potentials improves the discussion on the holographic modeling of a hadronic spectrum at zero temperature, where the masses play a central role.

\begin{acknowledgments}
The authors thank the Referee for their enlightening comments. M. A. Martin Contreras acknowledges the financial support provided by the National Natural Science Foundation of China (NSFC) under grant No 12350410371. S. M. Diles thanks to the Conselho Nacional de Desenvolvimento Científico e Tecnológico (CNPq), Brazil, Grant No. 406875/2023-5.
\end{acknowledgments}
\bibliography{apssamp}

\providecommand{\noopsort}[1]{}\providecommand{\singleletter}[1]{#1}%
\begin{thebibliography}{51}%
\makeatletter
\providecommand \@ifxundefined [1]{%
 \@ifx{#1\undefined}
}%
\providecommand \@ifnum [1]{%
 \ifnum #1\expandafter \@firstoftwo
 \else \expandafter \@secondoftwo
 \fi
}%
\providecommand \@ifx [1]{%
 \ifx #1\expandafter \@firstoftwo
 \else \expandafter \@secondoftwo
 \fi
}%
\providecommand \natexlab [1]{#1}%
\providecommand \enquote  [1]{``#1''}%
\providecommand \bibnamefont  [1]{#1}%
\providecommand \bibfnamefont [1]{#1}%
\providecommand \citenamefont [1]{#1}%
\providecommand \href@noop [0]{\@secondoftwo}%
\providecommand \href [0]{\begingroup \@sanitize@url \@href}%
\providecommand \@href[1]{\@@startlink{#1}\@@href}%
\providecommand \@@href[1]{\endgroup#1\@@endlink}%
\providecommand \@sanitize@url [0]{\catcode `\\12\catcode `\$12\catcode `\&12\catcode `\#12\catcode `\^12\catcode `\_12\catcode `\%12\relax}%
\providecommand \@@startlink[1]{}%
\providecommand \@@endlink[0]{}%
\providecommand \url  [0]{\begingroup\@sanitize@url \@url }%
\providecommand \@url [1]{\endgroup\@href {#1}{\urlprefix }}%
\providecommand \urlprefix  [0]{URL }%
\providecommand \Eprint [0]{\href }%
\providecommand \doibase [0]{https://doi.org/}%
\providecommand \selectlanguage [0]{\@gobble}%
\providecommand \bibinfo  [0]{\@secondoftwo}%
\providecommand \bibfield  [0]{\@secondoftwo}%
\providecommand \translation [1]{[#1]}%
\providecommand \BibitemOpen [0]{}%
\providecommand \bibitemStop [0]{}%
\providecommand \bibitemNoStop [0]{.\EOS\space}%
\providecommand \EOS [0]{\spacefactor3000\relax}%
\providecommand \BibitemShut  [1]{\csname bibitem#1\endcsname}%
\let\auto@bib@innerbib\@empty
\bibitem [{\citenamefont {Brodsky}\ and\ \citenamefont {de~Teramond}(2008)}]{Brodsky:2007hb}%
  \BibitemOpen
  \bibfield  {author} {\bibinfo {author} {\bibfnamefont {S.~J.}\ \bibnamefont {Brodsky}}\ and\ \bibinfo {author} {\bibfnamefont {G.~F.}\ \bibnamefont {de~Teramond}},\ }\bibfield  {title} {\bibinfo {title} {{Light-Front Dynamics and AdS/QCD Correspondence: The Pion Form Factor in the Space- and Time-Like Regions}},\ }\href {https://doi.org/10.1103/PhysRevD.77.056007} {\bibfield  {journal} {\bibinfo  {journal} {Phys. Rev. D}\ }\textbf {\bibinfo {volume} {77}},\ \bibinfo {pages} {056007} (\bibinfo {year} {2008})},\ \Eprint {https://arxiv.org/abs/0707.3859} {arXiv:0707.3859 [hep-ph]} \BibitemShut {NoStop}%
\bibitem [{\citenamefont {Abidin}\ and\ \citenamefont {Carlson}(2009)}]{Abidin:2009hr}%
  \BibitemOpen
  \bibfield  {author} {\bibinfo {author} {\bibfnamefont {Z.}~\bibnamefont {Abidin}}\ and\ \bibinfo {author} {\bibfnamefont {C.~E.}\ \bibnamefont {Carlson}},\ }\bibfield  {title} {\bibinfo {title} {{Nucleon electromagnetic and gravitational form factors from holography}},\ }\href {https://doi.org/10.1103/PhysRevD.79.115003} {\bibfield  {journal} {\bibinfo  {journal} {Phys. Rev. D}\ }\textbf {\bibinfo {volume} {79}},\ \bibinfo {pages} {115003} (\bibinfo {year} {2009})},\ \Eprint {https://arxiv.org/abs/0903.4818} {arXiv:0903.4818 [hep-ph]} \BibitemShut {NoStop}%
\bibitem [{\citenamefont {Martin~Contreras}\ \emph {et~al.}(2022{\natexlab{a}})\citenamefont {Martin~Contreras}, \citenamefont {Folco~Capossoli}, \citenamefont {Li}, \citenamefont {Vega},\ and\ \citenamefont {Boschi-Filho}}]{MartinContreras:2021yfz}%
  \BibitemOpen
  \bibfield  {author} {\bibinfo {author} {\bibfnamefont {M.~A.}\ \bibnamefont {Martin~Contreras}}, \bibinfo {author} {\bibfnamefont {E.}~\bibnamefont {Folco~Capossoli}}, \bibinfo {author} {\bibfnamefont {D.}~\bibnamefont {Li}}, \bibinfo {author} {\bibfnamefont {A.}~\bibnamefont {Vega}},\ and\ \bibinfo {author} {\bibfnamefont {H.}~\bibnamefont {Boschi-Filho}},\ }\bibfield  {title} {\bibinfo {title} {{Pion form factor from an AdS deformed background}},\ }\href {https://doi.org/10.1016/j.nuclphysb.2022.115726} {\bibfield  {journal} {\bibinfo  {journal} {Nucl. Phys. B}\ }\textbf {\bibinfo {volume} {977}},\ \bibinfo {pages} {115726} (\bibinfo {year} {2022}{\natexlab{a}})},\ \Eprint {https://arxiv.org/abs/2104.04640} {arXiv:2104.04640 [hep-ph]} \BibitemShut {NoStop}%
\bibitem [{\citenamefont {Folco~Capossoli}\ \emph {et~al.}(2020)\citenamefont {Folco~Capossoli}, \citenamefont {Mart\'\i{}n~Contreras}, \citenamefont {Li}, \citenamefont {Vega},\ and\ \citenamefont {Boschi-Filho}}]{FolcoCapossoli:2020pks}%
  \BibitemOpen
  \bibfield  {author} {\bibinfo {author} {\bibfnamefont {E.}~\bibnamefont {Folco~Capossoli}}, \bibinfo {author} {\bibfnamefont {M.~A.}\ \bibnamefont {Mart\'\i{}n~Contreras}}, \bibinfo {author} {\bibfnamefont {D.}~\bibnamefont {Li}}, \bibinfo {author} {\bibfnamefont {A.}~\bibnamefont {Vega}},\ and\ \bibinfo {author} {\bibfnamefont {H.}~\bibnamefont {Boschi-Filho}},\ }\bibfield  {title} {\bibinfo {title} {{Proton structure functions from an AdS/QCD model with a deformed background}},\ }\href {https://doi.org/10.1103/PhysRevD.102.086004} {\bibfield  {journal} {\bibinfo  {journal} {Phys. Rev. D}\ }\textbf {\bibinfo {volume} {102}},\ \bibinfo {pages} {086004} (\bibinfo {year} {2020})},\ \Eprint {https://arxiv.org/abs/2007.09283} {arXiv:2007.09283 [hep-ph]} \BibitemShut {NoStop}%
\bibitem [{\citenamefont {Braga}\ \emph {et~al.}(2016)\citenamefont {Braga}, \citenamefont {Martin~Contreras},\ and\ \citenamefont {Diles}}]{Braga:2015jca}%
  \BibitemOpen
  \bibfield  {author} {\bibinfo {author} {\bibfnamefont {N.~R.~F.}\ \bibnamefont {Braga}}, \bibinfo {author} {\bibfnamefont {M.~A.}\ \bibnamefont {Martin~Contreras}},\ and\ \bibinfo {author} {\bibfnamefont {S.}~\bibnamefont {Diles}},\ }\bibfield  {title} {\bibinfo {title} {{Decay constants in soft wall AdS/QCD revisited}},\ }\href {https://doi.org/10.1016/j.physletb.2016.10.046} {\bibfield  {journal} {\bibinfo  {journal} {Phys. Lett. B}\ }\textbf {\bibinfo {volume} {763}},\ \bibinfo {pages} {203} (\bibinfo {year} {2016})},\ \Eprint {https://arxiv.org/abs/1507.04708} {arXiv:1507.04708 [hep-th]} \BibitemShut {NoStop}%
\bibitem [{\citenamefont {Martin~Contreras}\ and\ \citenamefont {Vega}(2020{\natexlab{a}})}]{MartinContreras:2019kah}%
  \BibitemOpen
  \bibfield  {author} {\bibinfo {author} {\bibfnamefont {M.~A.}\ \bibnamefont {Martin~Contreras}}\ and\ \bibinfo {author} {\bibfnamefont {A.}~\bibnamefont {Vega}},\ }\bibfield  {title} {\bibinfo {title} {{Different approach to decay constants in AdS/QCD models}},\ }\href {https://doi.org/10.1103/PhysRevD.101.046009} {\bibfield  {journal} {\bibinfo  {journal} {Phys. Rev. D}\ }\textbf {\bibinfo {volume} {101}},\ \bibinfo {pages} {046009} (\bibinfo {year} {2020}{\natexlab{a}})},\ \Eprint {https://arxiv.org/abs/1910.10922} {arXiv:1910.10922 [hep-th]} \BibitemShut {NoStop}%
\bibitem [{\citenamefont {Vega}\ and\ \citenamefont {Cabrera}(2016)}]{Vega:2016gip}%
  \BibitemOpen
  \bibfield  {author} {\bibinfo {author} {\bibfnamefont {A.}~\bibnamefont {Vega}}\ and\ \bibinfo {author} {\bibfnamefont {P.}~\bibnamefont {Cabrera}},\ }\bibfield  {title} {\bibinfo {title} {{Family of dilatons and metrics for AdS/QCD models}},\ }\href {https://doi.org/10.1103/PhysRevD.93.114026} {\bibfield  {journal} {\bibinfo  {journal} {Phys. Rev. D}\ }\textbf {\bibinfo {volume} {93}},\ \bibinfo {pages} {114026} (\bibinfo {year} {2016})},\ \Eprint {https://arxiv.org/abs/1601.05999} {arXiv:1601.05999 [hep-ph]} \BibitemShut {NoStop}%
\bibitem [{\citenamefont {Afonin}\ and\ \citenamefont {Katanaeva}(2018)}]{Afonin:2018era}%
  \BibitemOpen
  \bibfield  {author} {\bibinfo {author} {\bibfnamefont {S.~S.}\ \bibnamefont {Afonin}}\ and\ \bibinfo {author} {\bibfnamefont {A.~D.}\ \bibnamefont {Katanaeva}},\ }\bibfield  {title} {\bibinfo {title} {{Glueballs and deconfinement temperature in AdS/QCD}},\ }\href {https://doi.org/10.1103/PhysRevD.98.114027} {\bibfield  {journal} {\bibinfo  {journal} {Phys. Rev. D}\ }\textbf {\bibinfo {volume} {98}},\ \bibinfo {pages} {114027} (\bibinfo {year} {2018})},\ \Eprint {https://arxiv.org/abs/1809.07730} {arXiv:1809.07730 [hep-ph]} \BibitemShut {NoStop}%
\bibitem [{\citenamefont {Martin~Contreras}\ and\ \citenamefont {Vega}(2023)}]{MartinContreras:2023oqs}%
  \BibitemOpen
  \bibfield  {author} {\bibinfo {author} {\bibfnamefont {M.~A.}\ \bibnamefont {Martin~Contreras}}\ and\ \bibinfo {author} {\bibfnamefont {A.}~\bibnamefont {Vega}},\ }\bibfield  {title} {\bibinfo {title} {{Holographic stability for non-qq\textasciimacron{} candidates}},\ }\href {https://doi.org/10.1103/PhysRevD.108.126024} {\bibfield  {journal} {\bibinfo  {journal} {Phys. Rev. D}\ }\textbf {\bibinfo {volume} {108}},\ \bibinfo {pages} {126024} (\bibinfo {year} {2023})},\ \Eprint {https://arxiv.org/abs/2309.02905} {arXiv:2309.02905 [hep-ph]} \BibitemShut {NoStop}%
\bibitem [{\citenamefont {Darboux}(1882)}]{darboux1882proposition}%
  \BibitemOpen
  \bibfield  {author} {\bibinfo {author} {\bibfnamefont {G.}~\bibnamefont {Darboux}},\ }\bibfield  {title} {\bibinfo {title} {Sur une proposition relative aux {\'e}quations li{\'e}aires},\ }\href@noop {} {\bibfield  {journal} {\bibinfo  {journal} {CR Acad. Sci. Paris}\ }\textbf {\bibinfo {volume} {94}},\ \bibinfo {pages} {1456} (\bibinfo {year} {1882})}\BibitemShut {NoStop}%
\bibitem [{\citenamefont {Abraham}\ and\ \citenamefont {Moses}(1980)}]{abraham1980changes}%
  \BibitemOpen
  \bibfield  {author} {\bibinfo {author} {\bibfnamefont {P.}~\bibnamefont {Abraham}}\ and\ \bibinfo {author} {\bibfnamefont {H.}~\bibnamefont {Moses}},\ }\bibfield  {title} {\bibinfo {title} {Changes in potentials due to changes in the point spectrum: anharmonic oscillators with exact solutions},\ }\href@noop {} {\bibfield  {journal} {\bibinfo  {journal} {Physical Review A}\ }\textbf {\bibinfo {volume} {22}},\ \bibinfo {pages} {1333} (\bibinfo {year} {1980})}\BibitemShut {NoStop}%
\bibitem [{\citenamefont {Mielnik}(1984)}]{mielnik1984factorization}%
  \BibitemOpen
  \bibfield  {author} {\bibinfo {author} {\bibfnamefont {B.}~\bibnamefont {Mielnik}},\ }\bibfield  {title} {\bibinfo {title} {Factorization method and new potentials with the oscillator spectrum},\ }\href@noop {} {\bibfield  {journal} {\bibinfo  {journal} {Journal of mathematical physics}\ }\textbf {\bibinfo {volume} {25}},\ \bibinfo {pages} {3387} (\bibinfo {year} {1984})}\BibitemShut {NoStop}%
\bibitem [{\citenamefont {Witten}(1981)}]{Witten:1981nf}%
  \BibitemOpen
  \bibfield  {author} {\bibinfo {author} {\bibfnamefont {E.}~\bibnamefont {Witten}},\ }\bibfield  {title} {\bibinfo {title} {{Dynamical Breaking of Supersymmetry}},\ }\href {https://doi.org/10.1016/0550-3213(81)90006-7} {\bibfield  {journal} {\bibinfo  {journal} {Nucl. Phys. B}\ }\textbf {\bibinfo {volume} {188}},\ \bibinfo {pages} {513} (\bibinfo {year} {1981})}\BibitemShut {NoStop}%
\bibitem [{\citenamefont {Cooper}\ \emph {et~al.}(1995)\citenamefont {Cooper}, \citenamefont {Khare},\ and\ \citenamefont {Sukhatme}}]{Cooper:1994eh}%
  \BibitemOpen
  \bibfield  {author} {\bibinfo {author} {\bibfnamefont {F.}~\bibnamefont {Cooper}}, \bibinfo {author} {\bibfnamefont {A.}~\bibnamefont {Khare}},\ and\ \bibinfo {author} {\bibfnamefont {U.}~\bibnamefont {Sukhatme}},\ }\bibfield  {title} {\bibinfo {title} {{Supersymmetry and quantum mechanics}},\ }\href {https://doi.org/10.1016/0370-1573(94)00080-M} {\bibfield  {journal} {\bibinfo  {journal} {Phys. Rept.}\ }\textbf {\bibinfo {volume} {251}},\ \bibinfo {pages} {267} (\bibinfo {year} {1995})},\ \Eprint {https://arxiv.org/abs/hep-th/9405029} {arXiv:hep-th/9405029} \BibitemShut {NoStop}%
\bibitem [{\citenamefont {Khare}\ and\ \citenamefont {Sukhatme}(1989)}]{Khare:1988is}%
  \BibitemOpen
  \bibfield  {author} {\bibinfo {author} {\bibfnamefont {A.}~\bibnamefont {Khare}}\ and\ \bibinfo {author} {\bibfnamefont {U.}~\bibnamefont {Sukhatme}},\ }\bibfield  {title} {\bibinfo {title} {{Phase Equivalent Potentials Obtained From Supersymmetry}},\ }\href {https://doi.org/10.1088/0305-4470/22/14/031} {\bibfield  {journal} {\bibinfo  {journal} {J. Phys. A}\ }\textbf {\bibinfo {volume} {22}},\ \bibinfo {pages} {2847} (\bibinfo {year} {1989})}\BibitemShut {NoStop}%
\bibitem [{\citenamefont {Aharony}\ \emph {et~al.}(2000)\citenamefont {Aharony}, \citenamefont {Gubser}, \citenamefont {Maldacena}, \citenamefont {Ooguri},\ and\ \citenamefont {Oz}}]{Aharony:1999ti}%
  \BibitemOpen
  \bibfield  {author} {\bibinfo {author} {\bibfnamefont {O.}~\bibnamefont {Aharony}}, \bibinfo {author} {\bibfnamefont {S.~S.}\ \bibnamefont {Gubser}}, \bibinfo {author} {\bibfnamefont {J.~M.}\ \bibnamefont {Maldacena}}, \bibinfo {author} {\bibfnamefont {H.}~\bibnamefont {Ooguri}},\ and\ \bibinfo {author} {\bibfnamefont {Y.}~\bibnamefont {Oz}},\ }\bibfield  {title} {\bibinfo {title} {{Large N field theories, string theory and gravity}},\ }\href {https://doi.org/10.1016/S0370-1573(99)00083-6} {\bibfield  {journal} {\bibinfo  {journal} {Phys. Rept.}\ }\textbf {\bibinfo {volume} {323}},\ \bibinfo {pages} {183} (\bibinfo {year} {2000})},\ \Eprint {https://arxiv.org/abs/hep-th/9905111} {arXiv:hep-th/9905111} \BibitemShut {NoStop}%
\bibitem [{\citenamefont {Gleiser}\ and\ \citenamefont {Stamatopoulos}(2012{\natexlab{a}})}]{Gleiser:2011di}%
  \BibitemOpen
  \bibfield  {author} {\bibinfo {author} {\bibfnamefont {M.}~\bibnamefont {Gleiser}}\ and\ \bibinfo {author} {\bibfnamefont {N.}~\bibnamefont {Stamatopoulos}},\ }\bibfield  {title} {\bibinfo {title} {{Entropic Measure for Localized Energy Configurations: Kinks, Bounces, and Bubbles}},\ }\href {https://doi.org/10.1016/j.physletb.2012.05.064} {\bibfield  {journal} {\bibinfo  {journal} {Phys. Lett. B}\ }\textbf {\bibinfo {volume} {713}},\ \bibinfo {pages} {304} (\bibinfo {year} {2012}{\natexlab{a}})},\ \Eprint {https://arxiv.org/abs/1111.5597} {arXiv:1111.5597 [hep-th]} \BibitemShut {NoStop}%
\bibitem [{\citenamefont {Gleiser}\ and\ \citenamefont {Stamatopoulos}(2012{\natexlab{b}})}]{Gleiser:2012tu}%
  \BibitemOpen
  \bibfield  {author} {\bibinfo {author} {\bibfnamefont {M.}~\bibnamefont {Gleiser}}\ and\ \bibinfo {author} {\bibfnamefont {N.}~\bibnamefont {Stamatopoulos}},\ }\bibfield  {title} {\bibinfo {title} {{Information Content of Spontaneous Symmetry Breaking}},\ }\href {https://doi.org/10.1103/PhysRevD.86.045004} {\bibfield  {journal} {\bibinfo  {journal} {Phys. Rev. D}\ }\textbf {\bibinfo {volume} {86}},\ \bibinfo {pages} {045004} (\bibinfo {year} {2012}{\natexlab{b}})},\ \Eprint {https://arxiv.org/abs/1205.3061} {arXiv:1205.3061 [hep-th]} \BibitemShut {NoStop}%
\bibitem [{\citenamefont {Bernardini}\ \emph {et~al.}(2017)\citenamefont {Bernardini}, \citenamefont {Braga},\ and\ \citenamefont {da~Rocha}}]{Bernardini:2016qit}%
  \BibitemOpen
  \bibfield  {author} {\bibinfo {author} {\bibfnamefont {A.~E.}\ \bibnamefont {Bernardini}}, \bibinfo {author} {\bibfnamefont {N.~R.~F.}\ \bibnamefont {Braga}},\ and\ \bibinfo {author} {\bibfnamefont {R.}~\bibnamefont {da~Rocha}},\ }\bibfield  {title} {\bibinfo {title} {{Configurational entropy of glueball states}},\ }\href {https://doi.org/10.1016/j.physletb.2016.12.007} {\bibfield  {journal} {\bibinfo  {journal} {Phys. Lett. B}\ }\textbf {\bibinfo {volume} {765}},\ \bibinfo {pages} {81} (\bibinfo {year} {2017})},\ \Eprint {https://arxiv.org/abs/1609.01258} {arXiv:1609.01258 [hep-th]} \BibitemShut {NoStop}%
\bibitem [{\citenamefont {Braga}\ and\ \citenamefont {da~Rocha}(2017)}]{Braga:2016wzx}%
  \BibitemOpen
  \bibfield  {author} {\bibinfo {author} {\bibfnamefont {N.~R.~F.}\ \bibnamefont {Braga}}\ and\ \bibinfo {author} {\bibfnamefont {R.}~\bibnamefont {da~Rocha}},\ }\bibfield  {title} {\bibinfo {title} {{Configurational entropy of anti-de Sitter black holes}},\ }\href {https://doi.org/10.1016/j.physletb.2017.02.031} {\bibfield  {journal} {\bibinfo  {journal} {Phys. Lett. B}\ }\textbf {\bibinfo {volume} {767}},\ \bibinfo {pages} {386} (\bibinfo {year} {2017})},\ \Eprint {https://arxiv.org/abs/1612.03289} {arXiv:1612.03289 [hep-th]} \BibitemShut {NoStop}%
\bibitem [{\citenamefont {Bernardini}\ and\ \citenamefont {da~Rocha}(2016)}]{Bernardini:2016hvx}%
  \BibitemOpen
  \bibfield  {author} {\bibinfo {author} {\bibfnamefont {A.~E.}\ \bibnamefont {Bernardini}}\ and\ \bibinfo {author} {\bibfnamefont {R.}~\bibnamefont {da~Rocha}},\ }\bibfield  {title} {\bibinfo {title} {{Entropic information of dynamical AdS/QCD holographic models}},\ }\href {https://doi.org/10.1016/j.physletb.2016.09.023} {\bibfield  {journal} {\bibinfo  {journal} {Phys. Lett. B}\ }\textbf {\bibinfo {volume} {762}},\ \bibinfo {pages} {107} (\bibinfo {year} {2016})},\ \Eprint {https://arxiv.org/abs/1605.00294} {arXiv:1605.00294 [hep-th]} \BibitemShut {NoStop}%
\bibitem [{\citenamefont {Braga}\ and\ \citenamefont {da~Rocha}(2018)}]{Braga:2017fsb}%
  \BibitemOpen
  \bibfield  {author} {\bibinfo {author} {\bibfnamefont {N.~R.~F.}\ \bibnamefont {Braga}}\ and\ \bibinfo {author} {\bibfnamefont {R.~a.}\ \bibnamefont {da~Rocha}},\ }\bibfield  {title} {\bibinfo {title} {{AdS/QCD duality and the quarkonia holographic information entropy}},\ }\href {https://doi.org/10.1016/j.physletb.2017.11.034} {\bibfield  {journal} {\bibinfo  {journal} {Phys. Lett. B}\ }\textbf {\bibinfo {volume} {776}},\ \bibinfo {pages} {78} (\bibinfo {year} {2018})},\ \Eprint {https://arxiv.org/abs/1710.07383} {arXiv:1710.07383 [hep-th]} \BibitemShut {NoStop}%
\bibitem [{\citenamefont {Colangelo}\ and\ \citenamefont {Loparco}(2019)}]{Colangelo:2018mrt}%
  \BibitemOpen
  \bibfield  {author} {\bibinfo {author} {\bibfnamefont {P.}~\bibnamefont {Colangelo}}\ and\ \bibinfo {author} {\bibfnamefont {F.}~\bibnamefont {Loparco}},\ }\bibfield  {title} {\bibinfo {title} {{Configurational Entropy can disentangle conventional hadrons from exotica}},\ }\href {https://doi.org/10.1016/j.physletb.2018.11.053} {\bibfield  {journal} {\bibinfo  {journal} {Phys. Lett. B}\ }\textbf {\bibinfo {volume} {788}},\ \bibinfo {pages} {500} (\bibinfo {year} {2019})},\ \Eprint {https://arxiv.org/abs/1811.05272} {arXiv:1811.05272 [hep-ph]} \BibitemShut {NoStop}%
\bibitem [{\citenamefont {Ferreira}\ and\ \citenamefont {da~Rocha}(2020{\natexlab{a}})}]{Ferreira:2019nkz}%
  \BibitemOpen
  \bibfield  {author} {\bibinfo {author} {\bibfnamefont {L.~F.}\ \bibnamefont {Ferreira}}\ and\ \bibinfo {author} {\bibfnamefont {R.}~\bibnamefont {da~Rocha}},\ }\bibfield  {title} {\bibinfo {title} {{Tensor mesons, AdS/QCD and information}},\ }\href {https://doi.org/10.1140/epjc/s10052-020-7978-7} {\bibfield  {journal} {\bibinfo  {journal} {Eur. Phys. J. C}\ }\textbf {\bibinfo {volume} {80}},\ \bibinfo {pages} {375} (\bibinfo {year} {2020}{\natexlab{a}})},\ \Eprint {https://arxiv.org/abs/1907.11809} {arXiv:1907.11809 [hep-th]} \BibitemShut {NoStop}%
\bibitem [{\citenamefont {Ferreira}\ and\ \citenamefont {da~Rocha}(2020{\natexlab{b}})}]{Ferreira:2020iry}%
  \BibitemOpen
  \bibfield  {author} {\bibinfo {author} {\bibfnamefont {L.~F.}\ \bibnamefont {Ferreira}}\ and\ \bibinfo {author} {\bibfnamefont {R.}~\bibnamefont {da~Rocha}},\ }\bibfield  {title} {\bibinfo {title} {{Nucleons and higher spin baryon resonances: An AdS/QCD configurational entropic incursion}},\ }\href {https://doi.org/10.1103/PhysRevD.101.106002} {\bibfield  {journal} {\bibinfo  {journal} {Phys. Rev. D}\ }\textbf {\bibinfo {volume} {101}},\ \bibinfo {pages} {106002} (\bibinfo {year} {2020}{\natexlab{b}})},\ \Eprint {https://arxiv.org/abs/2004.04551} {arXiv:2004.04551 [hep-th]} \BibitemShut {NoStop}%
\bibitem [{\citenamefont {da~Rocha}(2021)}]{daRocha:2021ntm}%
  \BibitemOpen
  \bibfield  {author} {\bibinfo {author} {\bibfnamefont {R.}~\bibnamefont {da~Rocha}},\ }\bibfield  {title} {\bibinfo {title} {{Information entropy in AdS/QCD: Mass spectroscopy of isovector mesons}},\ }\href {https://doi.org/10.1103/PhysRevD.103.106027} {\bibfield  {journal} {\bibinfo  {journal} {Phys. Rev. D}\ }\textbf {\bibinfo {volume} {103}},\ \bibinfo {pages} {106027} (\bibinfo {year} {2021})},\ \Eprint {https://arxiv.org/abs/2103.03924} {arXiv:2103.03924 [hep-ph]} \BibitemShut {NoStop}%
\bibitem [{\citenamefont {Braga}\ \emph {et~al.}(2018)\citenamefont {Braga}, \citenamefont {Ferreira},\ and\ \citenamefont {Da~Rocha}}]{Braga:2018fyc}%
  \BibitemOpen
  \bibfield  {author} {\bibinfo {author} {\bibfnamefont {N.~R.~F.}\ \bibnamefont {Braga}}, \bibinfo {author} {\bibfnamefont {L.~F.}\ \bibnamefont {Ferreira}},\ and\ \bibinfo {author} {\bibfnamefont {R.~a.}\ \bibnamefont {Da~Rocha}},\ }\bibfield  {title} {\bibinfo {title} {{Thermal dissociation of heavy mesons and configurational entropy}},\ }\href {https://doi.org/10.1016/j.physletb.2018.10.036} {\bibfield  {journal} {\bibinfo  {journal} {Phys. Lett. B}\ }\textbf {\bibinfo {volume} {787}},\ \bibinfo {pages} {16} (\bibinfo {year} {2018})},\ \Eprint {https://arxiv.org/abs/1808.10499} {arXiv:1808.10499 [hep-ph]} \BibitemShut {NoStop}%
\bibitem [{\citenamefont {Braga}\ and\ \citenamefont {da~Mata}(2020{\natexlab{a}})}]{Braga:2020hhs}%
  \BibitemOpen
  \bibfield  {author} {\bibinfo {author} {\bibfnamefont {N.~R.~F.}\ \bibnamefont {Braga}}\ and\ \bibinfo {author} {\bibfnamefont {R.}~\bibnamefont {da~Mata}},\ }\bibfield  {title} {\bibinfo {title} {{Configuration entropy description of charmonium dissociation under the influence of magnetic fields}},\ }\href {https://doi.org/10.1016/j.physletb.2020.135918} {\bibfield  {journal} {\bibinfo  {journal} {Phys. Lett. B}\ }\textbf {\bibinfo {volume} {811}},\ \bibinfo {pages} {135918} (\bibinfo {year} {2020}{\natexlab{a}})},\ \Eprint {https://arxiv.org/abs/2008.10457} {arXiv:2008.10457 [hep-th]} \BibitemShut {NoStop}%
\bibitem [{\citenamefont {Braga}\ and\ \citenamefont {da~Mata}(2020{\natexlab{b}})}]{Braga:2020myi}%
  \BibitemOpen
  \bibfield  {author} {\bibinfo {author} {\bibfnamefont {N.~R.~F.}\ \bibnamefont {Braga}}\ and\ \bibinfo {author} {\bibfnamefont {R.}~\bibnamefont {da~Mata}},\ }\bibfield  {title} {\bibinfo {title} {{Configuration entropy for quarkonium in a finite density plasma}},\ }\href {https://doi.org/10.1103/PhysRevD.101.105016} {\bibfield  {journal} {\bibinfo  {journal} {Phys. Rev. D}\ }\textbf {\bibinfo {volume} {101}},\ \bibinfo {pages} {105016} (\bibinfo {year} {2020}{\natexlab{b}})},\ \Eprint {https://arxiv.org/abs/2002.09413} {arXiv:2002.09413 [hep-th]} \BibitemShut {NoStop}%
\bibitem [{\citenamefont {Braga}\ and\ \citenamefont {Junqueira}(2021)}]{Braga:2020opg}%
  \BibitemOpen
  \bibfield  {author} {\bibinfo {author} {\bibfnamefont {N.~R.~F.}\ \bibnamefont {Braga}}\ and\ \bibinfo {author} {\bibfnamefont {O.~C.}\ \bibnamefont {Junqueira}},\ }\bibfield  {title} {\bibinfo {title} {{Configuration entropy and confinement/deconfinement transiton in holographic QCD}},\ }\href {https://doi.org/10.1016/j.physletb.2021.136082} {\bibfield  {journal} {\bibinfo  {journal} {Phys. Lett. B}\ }\textbf {\bibinfo {volume} {814}},\ \bibinfo {pages} {136082} (\bibinfo {year} {2021})},\ \Eprint {https://arxiv.org/abs/2010.00714} {arXiv:2010.00714 [hep-th]} \BibitemShut {NoStop}%
\bibitem [{\citenamefont {Martin~Contreras}\ \emph {et~al.}(2022{\natexlab{b}})\citenamefont {Martin~Contreras}, \citenamefont {Vega},\ and\ \citenamefont {Diles}}]{MartinContreras:2022lxl}%
  \BibitemOpen
  \bibfield  {author} {\bibinfo {author} {\bibfnamefont {M.~A.}\ \bibnamefont {Martin~Contreras}}, \bibinfo {author} {\bibfnamefont {A.}~\bibnamefont {Vega}},\ and\ \bibinfo {author} {\bibfnamefont {S.}~\bibnamefont {Diles}},\ }\bibfield  {title} {\bibinfo {title} {{A holographic bottom-up description of light nuclide spectroscopy and stability}},\ }\href {https://doi.org/10.1016/j.physletb.2022.137551} {\bibfield  {journal} {\bibinfo  {journal} {Phys. Lett. B}\ }\textbf {\bibinfo {volume} {835}},\ \bibinfo {pages} {137551} (\bibinfo {year} {2022}{\natexlab{b}})},\ \Eprint {https://arxiv.org/abs/2206.01834} {arXiv:2206.01834 [hep-ph]} \BibitemShut {NoStop}%
\bibitem [{\citenamefont {Boschi-Filho}\ and\ \citenamefont {Braga}(2004)}]{Boschi-Filho:2002wdj}%
  \BibitemOpen
  \bibfield  {author} {\bibinfo {author} {\bibfnamefont {H.}~\bibnamefont {Boschi-Filho}}\ and\ \bibinfo {author} {\bibfnamefont {N.~R.~F.}\ \bibnamefont {Braga}},\ }\bibfield  {title} {\bibinfo {title} {{QCD / string holographic mapping and glueball mass spectrum}},\ }\href {https://doi.org/10.1140/epjc/s2003-01526-4} {\bibfield  {journal} {\bibinfo  {journal} {Eur. Phys. J. C}\ }\textbf {\bibinfo {volume} {32}},\ \bibinfo {pages} {529} (\bibinfo {year} {2004})},\ \Eprint {https://arxiv.org/abs/hep-th/0209080} {arXiv:hep-th/0209080} \BibitemShut {NoStop}%
\bibitem [{\citenamefont {Erlich}\ \emph {et~al.}(2005)\citenamefont {Erlich}, \citenamefont {Katz}, \citenamefont {Son},\ and\ \citenamefont {Stephanov}}]{Erlich:2005qh}%
  \BibitemOpen
  \bibfield  {author} {\bibinfo {author} {\bibfnamefont {J.}~\bibnamefont {Erlich}}, \bibinfo {author} {\bibfnamefont {E.}~\bibnamefont {Katz}}, \bibinfo {author} {\bibfnamefont {D.~T.}\ \bibnamefont {Son}},\ and\ \bibinfo {author} {\bibfnamefont {M.~A.}\ \bibnamefont {Stephanov}},\ }\bibfield  {title} {\bibinfo {title} {{QCD and a holographic model of hadrons}},\ }\href {https://doi.org/10.1103/PhysRevLett.95.261602} {\bibfield  {journal} {\bibinfo  {journal} {Phys. Rev. Lett.}\ }\textbf {\bibinfo {volume} {95}},\ \bibinfo {pages} {261602} (\bibinfo {year} {2005})},\ \Eprint {https://arxiv.org/abs/hep-ph/0501128} {arXiv:hep-ph/0501128} \BibitemShut {NoStop}%
\bibitem [{\citenamefont {Karch}\ \emph {et~al.}(2006)\citenamefont {Karch}, \citenamefont {Katz}, \citenamefont {Son},\ and\ \citenamefont {Stephanov}}]{Karch:2006pv}%
  \BibitemOpen
  \bibfield  {author} {\bibinfo {author} {\bibfnamefont {A.}~\bibnamefont {Karch}}, \bibinfo {author} {\bibfnamefont {E.}~\bibnamefont {Katz}}, \bibinfo {author} {\bibfnamefont {D.~T.}\ \bibnamefont {Son}},\ and\ \bibinfo {author} {\bibfnamefont {M.~A.}\ \bibnamefont {Stephanov}},\ }\bibfield  {title} {\bibinfo {title} {{Linear confinement and AdS/QCD}},\ }\href {https://doi.org/10.1103/PhysRevD.74.015005} {\bibfield  {journal} {\bibinfo  {journal} {Phys. Rev. D}\ }\textbf {\bibinfo {volume} {74}},\ \bibinfo {pages} {015005} (\bibinfo {year} {2006})},\ \Eprint {https://arxiv.org/abs/hep-ph/0602229} {arXiv:hep-ph/0602229} \BibitemShut {NoStop}%
\bibitem [{\citenamefont {Braga}\ \emph {et~al.}(2017)\citenamefont {Braga}, \citenamefont {Ferreira},\ and\ \citenamefont {Vega}}]{Braga:2017bml}%
  \BibitemOpen
  \bibfield  {author} {\bibinfo {author} {\bibfnamefont {N.~R.~F.}\ \bibnamefont {Braga}}, \bibinfo {author} {\bibfnamefont {L.~F.}\ \bibnamefont {Ferreira}},\ and\ \bibinfo {author} {\bibfnamefont {A.}~\bibnamefont {Vega}},\ }\bibfield  {title} {\bibinfo {title} {{Holographic model for charmonium dissociation}},\ }\href {https://doi.org/10.1016/j.physletb.2017.10.013} {\bibfield  {journal} {\bibinfo  {journal} {Phys. Lett. B}\ }\textbf {\bibinfo {volume} {774}},\ \bibinfo {pages} {476} (\bibinfo {year} {2017})},\ \Eprint {https://arxiv.org/abs/1709.05326} {arXiv:1709.05326 [hep-ph]} \BibitemShut {NoStop}%
\bibitem [{\citenamefont {Martin~Contreras}\ and\ \citenamefont {Vega}(2020{\natexlab{b}})}]{MartinContreras:2020cyg}%
  \BibitemOpen
  \bibfield  {author} {\bibinfo {author} {\bibfnamefont {M.~A.}\ \bibnamefont {Martin~Contreras}}\ and\ \bibinfo {author} {\bibfnamefont {A.}~\bibnamefont {Vega}},\ }\bibfield  {title} {\bibinfo {title} {{Nonlinear Regge trajectories with AdS/QCD}},\ }\href {https://doi.org/10.1103/PhysRevD.102.046007} {\bibfield  {journal} {\bibinfo  {journal} {Phys. Rev. D}\ }\textbf {\bibinfo {volume} {102}},\ \bibinfo {pages} {046007} (\bibinfo {year} {2020}{\natexlab{b}})},\ \Eprint {https://arxiv.org/abs/2004.10286} {arXiv:2004.10286 [hep-ph]} \BibitemShut {NoStop}%
\bibitem [{\citenamefont {Martin~Contreras}\ \emph {et~al.}(2021)\citenamefont {Martin~Contreras}, \citenamefont {Diles},\ and\ \citenamefont {Vega}}]{MartinContreras:2021bis}%
  \BibitemOpen
  \bibfield  {author} {\bibinfo {author} {\bibfnamefont {M.~A.}\ \bibnamefont {Martin~Contreras}}, \bibinfo {author} {\bibfnamefont {S.}~\bibnamefont {Diles}},\ and\ \bibinfo {author} {\bibfnamefont {A.}~\bibnamefont {Vega}},\ }\bibfield  {title} {\bibinfo {title} {{Heavy quarkonia spectroscopy at zero and finite temperature in bottom-up AdS/QCD}},\ }\href {https://doi.org/10.1103/PhysRevD.103.086008} {\bibfield  {journal} {\bibinfo  {journal} {Phys. Rev. D}\ }\textbf {\bibinfo {volume} {103}},\ \bibinfo {pages} {086008} (\bibinfo {year} {2021})},\ \Eprint {https://arxiv.org/abs/2101.06212} {arXiv:2101.06212 [hep-ph]} \BibitemShut {NoStop}%
\bibitem [{\citenamefont {Workman}\ and\ \citenamefont {Others}(2022)}]{Workman:2022ynf}%
  \BibitemOpen
  \bibfield  {author} {\bibinfo {author} {\bibfnamefont {R.~L.}\ \bibnamefont {Workman}}\ and\ \bibinfo {author} {\bibnamefont {Others}} (\bibinfo {collaboration} {Particle Data Group}),\ }\bibfield  {title} {\bibinfo {title} {{Review of Particle Physics}},\ }\href {https://doi.org/10.1093/ptep/ptac097} {\bibfield  {journal} {\bibinfo  {journal} {PTEP}\ }\textbf {\bibinfo {volume} {2022}},\ \bibinfo {pages} {083C01} (\bibinfo {year} {2022})}\BibitemShut {NoStop}%
\bibitem [{\citenamefont {Afonin}\ \emph {et~al.}(2004)\citenamefont {Afonin}, \citenamefont {Andrianov}, \citenamefont {Andrianov},\ and\ \citenamefont {Espriu}}]{Afonin:2004yb}%
  \BibitemOpen
  \bibfield  {author} {\bibinfo {author} {\bibfnamefont {S.~S.}\ \bibnamefont {Afonin}}, \bibinfo {author} {\bibfnamefont {A.~A.}\ \bibnamefont {Andrianov}}, \bibinfo {author} {\bibfnamefont {V.~A.}\ \bibnamefont {Andrianov}},\ and\ \bibinfo {author} {\bibfnamefont {D.}~\bibnamefont {Espriu}},\ }\bibfield  {title} {\bibinfo {title} {{Matching Regge theory to the OPE}},\ }\href {https://doi.org/10.1088/1126-6708/2004/04/039} {\bibfield  {journal} {\bibinfo  {journal} {JHEP}\ }\textbf {\bibinfo {volume} {04}},\ \bibinfo {pages} {039}},\ \Eprint {https://arxiv.org/abs/hep-ph/0403268} {arXiv:hep-ph/0403268} \BibitemShut {NoStop}%
\bibitem [{\citenamefont {Hong}\ \emph {et~al.}(2006)\citenamefont {Hong}, \citenamefont {Yoon},\ and\ \citenamefont {Strassler}}]{Hong:2004sa}%
  \BibitemOpen
  \bibfield  {author} {\bibinfo {author} {\bibfnamefont {S.}~\bibnamefont {Hong}}, \bibinfo {author} {\bibfnamefont {S.}~\bibnamefont {Yoon}},\ and\ \bibinfo {author} {\bibfnamefont {M.~J.}\ \bibnamefont {Strassler}},\ }\bibfield  {title} {\bibinfo {title} {{On the couplings of vector mesons in AdS / QCD}},\ }\href {https://doi.org/10.1088/1126-6708/2006/04/003} {\bibfield  {journal} {\bibinfo  {journal} {JHEP}\ }\textbf {\bibinfo {volume} {04}},\ \bibinfo {pages} {003}},\ \Eprint {https://arxiv.org/abs/hep-th/0409118} {arXiv:hep-th/0409118} \BibitemShut {NoStop}%
\bibitem [{\citenamefont {Afonin}(2010)}]{Afonin:2010fr}%
  \BibitemOpen
  \bibfield  {author} {\bibinfo {author} {\bibfnamefont {S.~S.}\ \bibnamefont {Afonin}},\ }\bibfield  {title} {\bibinfo {title} {{Holographic like models as a five-dimensional rewriting of large-Nc QCD}},\ }\href {https://doi.org/10.1142/S0217751X10051049} {\bibfield  {journal} {\bibinfo  {journal} {Int. J. Mod. Phys. A}\ }\textbf {\bibinfo {volume} {25}},\ \bibinfo {pages} {5683} (\bibinfo {year} {2010})},\ \Eprint {https://arxiv.org/abs/1001.3105} {arXiv:1001.3105 [hep-ph]} \BibitemShut {NoStop}%
\bibitem [{\citenamefont {Andreev}(2006)}]{Andreev:2006vy}%
  \BibitemOpen
  \bibfield  {author} {\bibinfo {author} {\bibfnamefont {O.}~\bibnamefont {Andreev}},\ }\bibfield  {title} {\bibinfo {title} {{1/q**2 corrections and gauge/string duality}},\ }\href {https://doi.org/10.1103/PhysRevD.73.107901} {\bibfield  {journal} {\bibinfo  {journal} {Phys. Rev. D}\ }\textbf {\bibinfo {volume} {73}},\ \bibinfo {pages} {107901} (\bibinfo {year} {2006})},\ \Eprint {https://arxiv.org/abs/hep-th/0603170} {arXiv:hep-th/0603170} \BibitemShut {NoStop}%
\bibitem [{\citenamefont {Chen}(2018)}]{Chen:2018nnr}%
  \BibitemOpen
  \bibfield  {author} {\bibinfo {author} {\bibfnamefont {J.-K.}\ \bibnamefont {Chen}},\ }\bibfield  {title} {\bibinfo {title} {{Regge trajectories for the mesons consisting of different quarks}},\ }\href {https://doi.org/10.1140/epjc/s10052-018-6134-0} {\bibfield  {journal} {\bibinfo  {journal} {Eur. Phys. J. C}\ }\textbf {\bibinfo {volume} {78}},\ \bibinfo {pages} {648} (\bibinfo {year} {2018})}\BibitemShut {NoStop}%
\bibitem [{\citenamefont {Afonin}\ and\ \citenamefont {Pusenkov}(2014)}]{Afonin:2014nya}%
  \BibitemOpen
  \bibfield  {author} {\bibinfo {author} {\bibfnamefont {S.~S.}\ \bibnamefont {Afonin}}\ and\ \bibinfo {author} {\bibfnamefont {I.~V.}\ \bibnamefont {Pusenkov}},\ }\bibfield  {title} {\bibinfo {title} {{Universal description of radially excited heavy and light vector mesons}},\ }\href {https://doi.org/10.1103/PhysRevD.90.094020} {\bibfield  {journal} {\bibinfo  {journal} {Phys. Rev. D}\ }\textbf {\bibinfo {volume} {90}},\ \bibinfo {pages} {094020} (\bibinfo {year} {2014})},\ \Eprint {https://arxiv.org/abs/1411.2390} {arXiv:1411.2390 [hep-ph]} \BibitemShut {NoStop}%
\bibitem [{\citenamefont {Brodsky}\ \emph {et~al.}(2016{\natexlab{a}})\citenamefont {Brodsky}, \citenamefont {de~T\'eramond}, \citenamefont {Dosch},\ and\ \citenamefont {Lorc\'e}}]{Brodsky:2016yod}%
  \BibitemOpen
  \bibfield  {author} {\bibinfo {author} {\bibfnamefont {S.~J.}\ \bibnamefont {Brodsky}}, \bibinfo {author} {\bibfnamefont {G.~F.}\ \bibnamefont {de~T\'eramond}}, \bibinfo {author} {\bibfnamefont {H.~G.}\ \bibnamefont {Dosch}},\ and\ \bibinfo {author} {\bibfnamefont {C.}~\bibnamefont {Lorc\'e}},\ }\bibfield  {title} {\bibinfo {title} {{Universal Effective Hadron Dynamics from Superconformal Algebra}},\ }\href {https://doi.org/10.1016/j.physletb.2016.05.068} {\bibfield  {journal} {\bibinfo  {journal} {Phys. Lett. B}\ }\textbf {\bibinfo {volume} {759}},\ \bibinfo {pages} {171} (\bibinfo {year} {2016}{\natexlab{a}})},\ \Eprint {https://arxiv.org/abs/1604.06746} {arXiv:1604.06746 [hep-ph]} \BibitemShut {NoStop}%
\bibitem [{\citenamefont {Brodsky}\ \emph {et~al.}(2016{\natexlab{b}})\citenamefont {Brodsky}, \citenamefont {de~T\'eramond}, \citenamefont {Dosch},\ and\ \citenamefont {Lorc\'e}}]{Brodsky:2016rvj}%
  \BibitemOpen
  \bibfield  {author} {\bibinfo {author} {\bibfnamefont {S.~J.}\ \bibnamefont {Brodsky}}, \bibinfo {author} {\bibfnamefont {G.~F.}\ \bibnamefont {de~T\'eramond}}, \bibinfo {author} {\bibfnamefont {H.~G.}\ \bibnamefont {Dosch}},\ and\ \bibinfo {author} {\bibfnamefont {C.}~\bibnamefont {Lorc\'e}},\ }\bibfield  {title} {\bibinfo {title} {{Meson/Baryon/Tetraquark Supersymmetry from Superconformal Algebra and Light-Front Holography}},\ }\href {https://doi.org/10.1142/S0217751X16300295} {\bibfield  {journal} {\bibinfo  {journal} {Int. J. Mod. Phys. A}\ }\textbf {\bibinfo {volume} {31}},\ \bibinfo {pages} {1630029} (\bibinfo {year} {2016}{\natexlab{b}})},\ \Eprint {https://arxiv.org/abs/1606.04638} {arXiv:1606.04638 [hep-ph]} \BibitemShut {NoStop}%
\bibitem [{\citenamefont {Nielsen}\ and\ \citenamefont {Brodsky}(2018)}]{Nielsen:2018uyn}%
  \BibitemOpen
  \bibfield  {author} {\bibinfo {author} {\bibfnamefont {M.}~\bibnamefont {Nielsen}}\ and\ \bibinfo {author} {\bibfnamefont {S.~J.}\ \bibnamefont {Brodsky}},\ }\bibfield  {title} {\bibinfo {title} {{Hadronic superpartners from a superconformal and supersymmetric algebra}},\ }\href {https://doi.org/10.1103/PhysRevD.97.114001} {\bibfield  {journal} {\bibinfo  {journal} {Phys. Rev. D}\ }\textbf {\bibinfo {volume} {97}},\ \bibinfo {pages} {114001} (\bibinfo {year} {2018})},\ \Eprint {https://arxiv.org/abs/1802.09652} {arXiv:1802.09652 [hep-ph]} \BibitemShut {NoStop}%
\bibitem [{\citenamefont {Nielsen}\ \emph {et~al.}(2018)\citenamefont {Nielsen}, \citenamefont {Brodsky}, \citenamefont {de~T\'eramond}, \citenamefont {Dosch}, \citenamefont {Navarra},\ and\ \citenamefont {Zou}}]{Nielsen:2018ytt}%
  \BibitemOpen
  \bibfield  {author} {\bibinfo {author} {\bibfnamefont {M.}~\bibnamefont {Nielsen}}, \bibinfo {author} {\bibfnamefont {S.~J.}\ \bibnamefont {Brodsky}}, \bibinfo {author} {\bibfnamefont {G.~F.}\ \bibnamefont {de~T\'eramond}}, \bibinfo {author} {\bibfnamefont {H.~G.}\ \bibnamefont {Dosch}}, \bibinfo {author} {\bibfnamefont {F.~S.}\ \bibnamefont {Navarra}},\ and\ \bibinfo {author} {\bibfnamefont {L.}~\bibnamefont {Zou}},\ }\bibfield  {title} {\bibinfo {title} {{Supersymmetry in the Double-Heavy Hadronic Spectrum}},\ }\href {https://doi.org/10.1103/PhysRevD.98.034002} {\bibfield  {journal} {\bibinfo  {journal} {Phys. Rev. D}\ }\textbf {\bibinfo {volume} {98}},\ \bibinfo {pages} {034002} (\bibinfo {year} {2018})},\ \Eprint {https://arxiv.org/abs/1805.11567} {arXiv:1805.11567 [hep-ph]} \BibitemShut {NoStop}%
\bibitem [{\citenamefont {da~Rocha}\ and\ \citenamefont {Silva}(2023)}]{daRocha:2023nsb}%
  \BibitemOpen
  \bibfield  {author} {\bibinfo {author} {\bibfnamefont {R.}~\bibnamefont {da~Rocha}}\ and\ \bibinfo {author} {\bibfnamefont {P.~H.~O.}\ \bibnamefont {Silva}},\ }\bibfield  {title} {\bibinfo {title} {{Configurational entropy and shape complexity of strange vector kaons in AdS/QCD}},\ }\href {https://doi.org/10.1140/epjp/s13360-023-04372-9} {\bibfield  {journal} {\bibinfo  {journal} {Eur. Phys. J. Plus}\ }\textbf {\bibinfo {volume} {138}},\ \bibinfo {pages} {729} (\bibinfo {year} {2023})},\ \Eprint {https://arxiv.org/abs/2302.05785} {arXiv:2302.05785 [hep-ph]} \BibitemShut {NoStop}%
\bibitem [{\citenamefont {Quigg}\ and\ \citenamefont {Rosner}(1978)}]{Quigg:1977wn}%
  \BibitemOpen
  \bibfield  {author} {\bibinfo {author} {\bibfnamefont {C.}~\bibnamefont {Quigg}}\ and\ \bibinfo {author} {\bibfnamefont {J.~L.}\ \bibnamefont {Rosner}},\ }\bibfield  {title} {\bibinfo {title} {{Semiclassical Sum Rules}},\ }\href {https://doi.org/10.1103/PhysRevD.17.2364} {\bibfield  {journal} {\bibinfo  {journal} {Phys. Rev. D}\ }\textbf {\bibinfo {volume} {17}},\ \bibinfo {pages} {2364} (\bibinfo {year} {1978})}\BibitemShut {NoStop}%
\bibitem [{\citenamefont {Lucha}\ \emph {et~al.}(1991)\citenamefont {Lucha}, \citenamefont {Schoberl},\ and\ \citenamefont {Gromes}}]{Lucha:1991vn}%
  \BibitemOpen
  \bibfield  {author} {\bibinfo {author} {\bibfnamefont {W.}~\bibnamefont {Lucha}}, \bibinfo {author} {\bibfnamefont {F.~F.}\ \bibnamefont {Schoberl}},\ and\ \bibinfo {author} {\bibfnamefont {D.}~\bibnamefont {Gromes}},\ }\bibfield  {title} {\bibinfo {title} {{Bound states of quarks}},\ }\href {https://doi.org/10.1016/0370-1573(91)90001-3} {\bibfield  {journal} {\bibinfo  {journal} {Phys. Rept.}\ }\textbf {\bibinfo {volume} {200}},\ \bibinfo {pages} {127} (\bibinfo {year} {1991})}\BibitemShut {NoStop}%
\end{thebibliography}%

\end{document}